# Molecular Pseudorotation in Phthalocyanines as a Tool for Magnetic Field Control at the Nanoscale

Raphael Wilhelmer, Matthias Diez, Johannes K. Krondorfer, and Andreas W. Hauser*

**ABSTRACT:** Metal phthalocyanines, a highly versatile class of aromatic, planar, macrocyclic molecules with a chelated central metal ion, are topical objects of ongoing research and particularly interesting due to their magnetic properties. However, while the current focus lies almost exclusively on spin-Zeeman-related effects, the high symmetry of the molecule and its circular shape suggests the exploitation of light-induced excitation of 2-fold degenerate vibrational states in order to generate, switch, and manipulate magnetic fields at the nanoscale. The underlying mechanism is a molecular pseudorotation that can be triggered by infrared pulses and gives rise to a quantized, small, but controllable magnetic dipole moment. We investigate the optical stimulation of vibrationally induced molecular magnetism and estimate changes in the magnetic shielding constants for confirmation by future experiments.

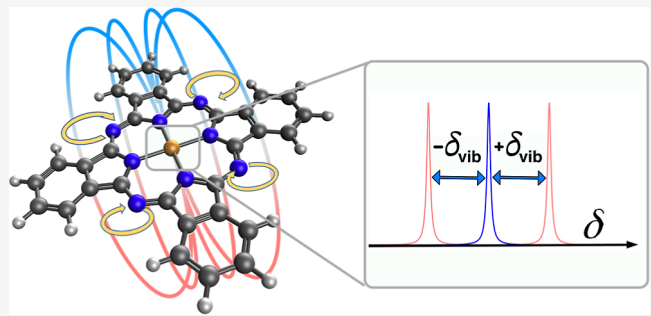

## 1. INTRODUCTION

The manipulation of organic molecules by magnetic fields is a highly active field of research with potential applications in molecular circuitry,[1,2] quantum computing,[3,4] optoelectronics,[5] and even the chiral separation of racemic mixtures via enantiospecific induced spin polarization.[6,7] Typically, paramagnetic states with nonzero spin multiplicity are the starting point for investigations of magnetic field sensitivity. When exposed to an external magnetic field, the degeneracy of the orientational spin quantum number is lifted, and changing the occupation of these magnetic sublevels is known to affect the chemical, optical and electronic properties of certain materials,[8−12] a concept that has been exploited by nature on numerous occasions in the form of biological sensors using earth's magnetic field for orientation.[13−15]

Nevertheless, on the road toward new multifunctional organic devices with integrated electronic, optical, and magnetic properties, studies on the diamagnetic properties of these materials have seen a very recent revival, in particular with respect to the phenomenon of ring currents of aromatic molecules.[16,17] Although the actual understanding of ring currents in chemical systems is still incomplete,[5,18−20] their consequences in nuclear magnetic resonance (NMR) spectroscopy are well studied.[21,22] A particularly interesting question is whether additional ring currents, driven by an external magnetic field, can also be used to control molecular features. Recently, a large step in this direction has been made by Kudisch et al., who demonstrated a change in the photophysical properties of $\pi$-stacked $H_2$−phthalocyanine molecules when exposed to a strong magnetic field.[5] The latter research is particularly noteworthy as most of the ongoing work on phthalocyanines is dedicated only to spin Zeeman coupling effects in the subclass of metal phthalocyanines. This abundant group of molecules has become one of the most studied organic materials for applications as catalysts, dyes, and coatings due to their optical, magnetic, and electronic properties. It has been studied in the bulk, as thin films, and in solution as well as on single molecules on various substrates.[23−25]

In this article, although closely related to the above in terms of its consequences, an entirely different handle on ring currents and magnetism of aromatic molecules is proposed by theory: a "concerted", two-dimensional molecular vibration, also known as pseudorotation. The latter term is commonly used to describe an intramolecular motion, resulting in a structure that appears to have been produced by a rotation of the entire initial molecule. Within the picture of vibrational eigenstates, motions of this type may be interpreted as simultaneous excitation of energetically degenerate vibrational modes. As is shown below, the linear combination of two degenerate, infrared (IR) active molecular vibrations, excited with a phase delay of $\pi/2$, can cause a pseudorotational motion, which gives rise to an electric dipole moment vector that rotates in the molecular plane and may be accompanied by a vibrationally induced intramolecular magnetic field. From an oversimplified but intuitive perspective, such a specific vibrational excitation may turn suitable molecules into a molecular analogue of Rowland's famous ring experiment,[26] as is shown in Figure 1. Regarding the magnetic field created by this contraption, its similarity to the field caused by a rotation of isolated charges on a ring, or a current flowing through a single loop of conducting wire, will be investigated in detail. We discuss the suitability of empirical and quantum-mechanical models in an attempt to describe electric dipole moment dynamics, an effective magnetic dipole moment, and the generation of an intramolecular vibrationally induced magnetic field.

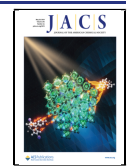

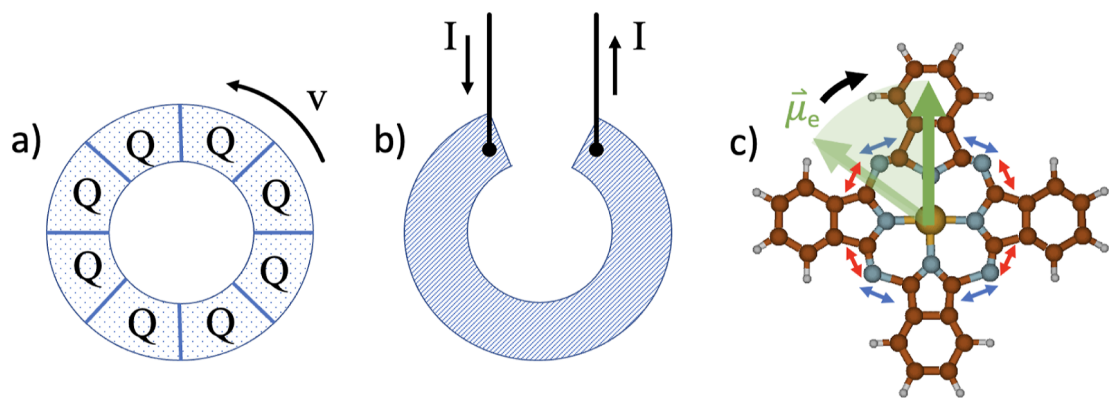

**Figure 1.** Three physical setups that create a magnetic field perpendicular to the paper plane: (a) Rowland's experiment of a rotating ring of charges, (b) a current $I$ passing through a loop of conductive material, and (c) a metal phthalocyanine, periodically deformed by two degenerate vibrations (red and blue, here mostly affecting outer N positions), giving rise to a rotation of an electric dipole moment (green) in the molecular plane.

While the combination of degenerate vibrational excitations with a specific phase relation has been a commonplace in theoretical molecular spectroscopy for decades,[27−29] e.g., to capture intramolecular rearrangements such as the name-giving Berry mechanism in highly symmetric molecules,[30] experimental control and exploitation has come into reach just recently by the advent of ultrashort laser spectroscopy.[31−33] However, most often, the focus is set on electronic degeneracies in aromatic molecules[34−37] and hence on spectroscopy in the ultraviolet/vis regime. Very closely related is the work of Juraschek et al.[38−40] on phonon-mediated dynamical multiferroicity, where a time-dependent polarization is proposed to induce magnetization in bulk materials.

Our article is structured as follows. In the first section, we provide an abbreviated review of the rotational $g$-factor, a dimensionless quantity connecting the angular momentum and magnetic moment, and its related spin-rotation coupling parameters. This is followed by the extension of the existing theories on the vibrational $g$-factor, in particular the semi-classical model of Moss and Perry,[41] to capture spin−vibration coupling as well, within a unified treatment for rotation and vibration. We investigate the coupling among molecular pseudorotation, an external magnetic field, and the nuclear spins within a given molecule. The underlying principle is the fact that pseudorotations involve the motion of certain atomic nuclei along closed loops and, as charge-carrying particles, they may generate magnetic fields.

In the main part, H$_2$Pc and CuPc molecules, two extreme representatives of the metal phthalocyanine complexes, are investigated in greater detail via density functional theory (DFT). The GFN2-$x$TB method, a tight binding ansatz of considerably reduced computational cost, is validated through direct comparison on these two molecules and applied to Mn−Pc, Fe−Pc, Co−Pc, and Ni−Pc in the Supporting Information. Given our interest in the possibility of photoexcitation of magnetic field-generating vibrational modes, we present IR spectra and select the strongest, IR-active planar pseudorotations for further study. Vibrational $g$-factors are then calculated, and changes in the dipole moment as well as the overall electron density distribution are presented as a function of the corresponding concerted vibrational motion. A particularly well-suited type of pseudorotation, typically located at about 1526 cm$^{-1}$ in all phtalocyanines under study, is selected for further investigation, and changes of the magnetic shielding constants in both directions, triggered by the optical excitation, are estimated for future experimental verification of vibrationally induced local magnetic fields.

## 2. METHODOLOGY

**2.1. Magnetic Effects due to Molecular Pseudorotation.** Cyclic motions in molecules or solids, where the positively charged nuclei are moving along closed orbits, give rise to the phenomenon of an "orbital" magnetic moment oriented perpendicular to the plane of the cyclic motion. To a large extent, this moment is screened by the atomically localized part of the electron density. In addition to the most obvious cyclic motion performed by a molecule, namely, its free rotation in space, molecular vibrations in the form of pseudorotations can also lead to circular motions of certain atoms. A simple and particularly insightful example featuring such a pseudorotation is the Na$_3$ molecule in its Jahn−Teller distorted $1^2E'$ ground state,[42,43] where spectroscopic consequences have been discussed in the greatest detail within the context of a "molecular" Berry phase.[44]

It is our aim to provide a suitable methodology for the analysis of magnetic fields arising from such combined vibrational excitations. While rotational $g$-factors and spin-rotation coupling matrices have been thoroughly discussed in the literature,[45−48] the investigation of vibrational and pseudorotational couplings has been restricted, so far, to $g$-factors only.[41] In this section, for the sake of a transparent discussion, we provide a short derivation of rotational and vibrational couplings. A more detailed derivation can be found in the Supporting Information.

Considering the effect of nuclear rotation and vibration, a complete description of nuclear motion is possible via center of mass coordinates, Euler angles, and vibrational coordinates evaluated at the minimum of a selected potential energy surface. After expressing the classical kinetic energy in these coordinates, one obtains a metric, which may be used to construct a corresponding quantum Hamiltonian for the entire molecular system by canonical quantization.[49,50] This procedure yields the total Hamiltonian

$$H = \sum_{i=0}^{N_e} \frac{\mathbf{p}_i^2}{2m_e} + V_c$$
$$+ \frac{1}{2}(\mathbf{J} - \mathbf{G} - \mathbf{L}_e)\Theta_{\text{eff}}^{-1}(\mathbf{J} - \mathbf{G} - \mathbf{L}_e)$$
$$+ \frac{1}{2}\sum_r^{3N_N-6} P_r^2 - \frac{\hbar^2}{8}\sum_{\alpha=1}^{N_n} \Theta_{\text{eff},\alpha\alpha}^{-1}, \quad (1)$$

where the first line corresponds to the electronic kinetic energy and all types of intramolecular Coulomb interactions, the second line corresponds to the rotational kinetic energy of the nuclear system in the respective coordinate frame, and the last line contains vibrational kinetic energy of the nuclei as well as a mass correction term, which is negligible. For small nuclear displacements, $\Theta_{\text{eff}}$ can be approximated by the nuclear inertia tensor $\Theta$ evaluated at the equilibrium geometry.

The nuclear rotational energy is relevant for vibrational and rotational coupling. Here, $\mathbf{J} = \mathbf{L}_e + \mathbf{L}_N + \mathbf{G}$ denotes the total angular momentum, consisting of an electronic contribution $\mathbf{L}_e = \sum_{i=0}^{N_e} \mathbf{L}_i$, a nuclear contribution $\mathbf{L}_N$, and a vibrational angular momentum $\mathbf{G}$. The latter is nonzero only for degenerate vibrational excitations and can be written as

$$\mathbf{G} = \sum_\alpha \mathbf{G}_\alpha = \sum_{\alpha,t} \zeta_{\alpha,t}(Q_{t_1}P_{t_2} - Q_{t_2}P_{t_1}) = \sum_t \zeta_t G_t \quad (2)$$

with $t$ indexing the doubly degenerate vibrational excitations and $Q_{t_1}$ and $Q_{t_2}$ denoting the respective normal coordinates. Furthermore, the scalar vibrational angular momentum $G_t$ for each degenerate pair of eigenmodes has been introduced as well as the contribution of a single nucleus, $\mathbf{G}_\alpha$. $\zeta_t$ denotes the so-called Coriolis coupling constant and $\zeta_{\alpha,t} = \mathbf{d}_{\alpha,t_1} \times \mathbf{d}_{\alpha,t_2}$ denotes the respective contribution of nucleus $\alpha$, with $\mathbf{d}_{\alpha,t_i}$ referring to the normalized, mass weighted displacement vector of the $i$-th vibrational mode.

We can perform first-order state correction with perturbation $-(\mathbf{J} - \mathbf{G})\Theta^{-1}\mathbf{L}_e$ for the electronic system to obtain

$$|0\rangle^1 = |0\rangle - \sum_{n\neq 0} \langle \mathbf{J} - \mathbf{G}\rangle\Theta^{-1}\frac{\langle n|\mathbf{L}_e|0\rangle}{E_0 - E_n}|n\rangle \quad (3)$$

The expression $\langle \mathbf{J} - \mathbf{G}\rangle$ denotes the expectation value with respect to the nuclear system. Usually, however, this term is treated classically. The electronic contribution to rotational and vibrational coupling parameters can then be calculated by evaluating the expectation value of a suitable interaction Hamiltonian in the perturbed state.

In order to obtain rotational and vibrational g-factors, we consider the paramagnetic interaction Hamiltonian with an external field $H_B^{\text{para}} = \left(\frac{e}{2m_e}\mathbf{L}_e - \sum_\alpha \frac{Z_\alpha}{2M_\alpha}\mathbf{L}_\alpha\right)\cdot\mathbf{B}$. Note that $\mathbf{L}_\alpha$ can be expressed as

$$\mathbf{L}_{\alpha,\text{com}} = \Theta^{(\alpha,\text{com})}\Theta^{-1}(\mathbf{J} - \mathbf{L}_e - \mathbf{G}) + \mathbf{G}_\alpha \quad (4)$$

with

$$\Theta^{(\alpha,\text{com})} = M_\alpha[(\mathbf{R}_\alpha - \mathbf{R}_{\text{com}})^2 \mathbf{1} - (\mathbf{R}_\alpha - \mathbf{R}_{\text{com}})(\mathbf{R}_\alpha - \mathbf{R}_{\text{com}})^T]$$

as the inertia tensor of nucleus $\alpha$ with respect to the center of mass. Using this and taking the expectation value in the perturbed state, this yields the rotationally and vibrationally induced magnetic moment $\boldsymbol{\mu}_{\text{mag}}^{\text{ind}} = g^{\text{rot}}\mathbf{J} - \sum_t g_t^{\text{vib}}G_t$ under the assumption of a $\Sigma$ electronic ground state. The rotational and vibrational g-factors are

$$g^{\text{rot}} = -\frac{2m_e}{e}\xi^{\text{para}}\Theta^{-1} + e\sum_\alpha \frac{Z_\alpha}{2M_\alpha}\Theta^{(\alpha,\text{com})}\Theta^{-1}$$

$$g_t^{\text{vib}} = \sum_k (g^{\text{rot}}\zeta_t) - \sum_\alpha \frac{eZ_\alpha(\zeta_{\alpha,t})}{2M_\alpha}. \quad (5)$$

with $\xi^{\text{para}}$ as the paramagnetic magnetizability. Equation 5 agrees with the known expressions of vibrational and rotational g-factors from the literature.[41,48] The magnetic moment, however, does not contain information about the magnetic field distribution within the molecule. Hence, to obtain measurable quantities, we calculate rotationally and vibrationally induced NMR splitting for each nucleus. The corresponding interaction Hamiltonian

$$H_{\text{SO}} = \sum_\alpha \frac{\mu_0}{4\pi}e\left[-\sum_i \frac{1}{m_e}\frac{\mathbf{L}_{i,\alpha}}{|\mathbf{R}_\alpha - \mathbf{r}_i|^3}\right.$$
$$\left.+ \sum_{\beta\neq\alpha} \frac{Z_\beta}{M_\beta}\frac{\mathbf{L}_{\beta,\alpha}}{|\mathbf{R}_\alpha - \mathbf{R}_\beta|^3}\right]\cdot\gamma_\alpha\mathbf{I}_\alpha \quad (6)$$

is a spin−orbit coupling Hamiltonian with $\mathbf{L}_{i,\alpha}$ as the electron angular momentum with respect to nucleus $\alpha$, $\mathbf{L}_{\beta,\alpha}$ as the nuclear angular momentum of nucleus $\beta$ with respect to nucleus $\alpha$, the nuclear gyromagnetic ratio $\gamma_\alpha$ and the nuclear spin operator $\mathbf{I}_\alpha$. Calculating the expectation value in the perturbed state yields the rotationally and vibrationally induced magnetic field at nucleus $\alpha$

$$\mathbf{B}_{\text{rot},\alpha}^{\text{ind}} = \frac{2m_e}{e}\sigma_\alpha^{\text{para}}\Theta^{-1}\mathbf{J} + \frac{\mu_0 e}{4\pi}\sum_\beta \frac{Z_\beta}{M_\beta}\frac{\Theta^{(\beta,\alpha)}\Theta^{-1}\mathbf{J}}{|\mathbf{R}_\alpha - \mathbf{R}_\beta|^3},$$

$$\mathbf{B}_{\text{vib},\alpha}^{\text{ind}} = -\frac{2m_e}{e}\sigma_\alpha^{\text{para}}\Theta^{-1}\mathbf{G} + \frac{\mu_0 e}{4\pi}$$
$$\sum_\beta \frac{Z_\beta}{M_\beta}\frac{\mathbf{G}_\beta - \Theta^{(\beta,\alpha)}\Theta^{-1}\mathbf{G}}{|\mathbf{R}_\alpha - \mathbf{R}_\beta|^3}, \quad (7)$$

with $\Theta^{(\beta,\alpha)}$ denoting the inertia tensor of nucleus $\beta$ with respect to nucleus $\alpha$ and $\sigma_\alpha^{\text{para}}$ denoting the paramagnetic shielding tensor at nucleus $\alpha$. For a rotationally induced magnetic field, this is consistent with ref 48 but generalizes the known result to the case of pseudorotational excitations.

**2.2. Pseudorotation in the Bulk.** In solid-state physics, the literature on an equivalent phenomenon speaks of 'dynamical multiferroicity' or a "phonon Zeeman effect".[38−40] Typically, magnetic moments in bulk structures are calculated from phonon modes and Born effective charge tensors. Significant effects are predicted for large charges in combination with small reduced masses, e.g., in hydride compounds such as CsH or CuH. Juraschek et al. investigated the possibility of inducing magnetization in spinless bulk materials by time-dependent electric polarization.[38] Interested in the effective magnetic moment $M$, they calculate the contributions of all atoms in the unit cell by linking their rotational movement to a "phonon" angular momentum $L$. Assuming that only two phonon modes 1 and 2 are contributing, they find

$$M = \gamma_{12} L \qquad (8)$$

with $\gamma_{12}$ denoting the gyromagnetic ratio, and derive the latter from the circular motions of effectively charged nuclei $\alpha$ via

$$\gamma_{12} = \sum_\alpha \gamma_\alpha \mathbf{d}_{\alpha,1} \times \mathbf{d}_{\alpha,2} \qquad (9)$$

using the same nomenclature as that introduced in Section 2.1 to distinguish the two orthogonal modes of a doubly degenerate vibration. The gyromagnetic ratio for each atom $\alpha$ is given by

$$\gamma_\alpha = \frac{e Z_\alpha^*}{2 M_\alpha} \qquad (10)$$

with $Z_\alpha^*$ denoting the Born effective charge obtained from DFT calculations. However, while this technique may provide useful predictions for macroscopic magnetic properties of bulk materials, a suitable description of magnetic field configurations on the subnanometer level can only be based on an analysis including electron density fluctuations upon vibrational excitation. This intrinsic limitation is addressed in Section 3.5.

**2.3. Optical Excitation of a Rotating Electric Dipole.** The dependence of the magnitude of the magnetic moment on the quadratic distance of a charge from its rotational axis, as can be seen, e.g., in the nuclear contribution to the $g$-factor in eq 5, clearly reminds of the classical definition of the magnetic moment

$$\mu = I \cdot A \qquad (11)$$

if we interpret the rotating charges as a classical, closed loop of current $I$ and area $A$. Assuming an opposite charge, but equal in magnitude, placed at the rotational center, it is tempting to imagine the charge dynamics during a pseudorotation as a rotating electric dipole vector. Indeed, as is shown below, the effective electric dipole moment of the phthalocyanines shows the behavior of a rotating vector during pseudorotation. However, the motion of the electric dipole moment $\mu_{el}$, an expectation value of the molecular system at a given geometry, does not directly translate into a magnetic moment for several reasons, as is discussed in the Results and Discussion section below.

Juraschek et al. further suggested to stimulate degenerate pairs of phonon modes via optical excitation, which would, in the case described above, lead to a magnetization that is proportional to the phonon frequency, the amplitude of the motion, and the phase relation between the two degenerate modes.[38] Within the harmonic approximation, each of them, denoted for convenience as the $Q_1$ and $Q_2$ modes, will drive a periodic modulation of the molecular electric dipole moment, e.g., in either the $x$ or $y$ direction, respectively. This modulation will also be harmonic in character. Assuming a simultaneous excitation of the two planar modes, albeit with a phase difference of $\pm \pi/2$, leads to the picture of the dipole moment vector $\vec{p}$ that rotates in that plane

$$\vec{p}(t) = (p_x(t), p_y(t)) = p \cdot (\sin(\omega t), \pm\cos(\omega t)). \qquad (12)$$

Although the picture of a rotating electric dipole moment is too crude when discussing actual magnetic field geometries at the nanoscale, it has proven to be a suitable model for dynamical studies of the macroscopic magnetization in the bulk as a response to pulsed IR light. Note that irrespective of its limited value for the actual characterization of molecular magnetic fields, the presence of an electric dipole moment upon vibrational excitation is a necessary prerequisite for the suggested optical stimulation of vibrationally induced magnetism.

**2.4. Computational Details.** Electronic structure calculations for the moderately sized phthalocyanine complexes are performed via DFT, employing the B97D functional, a 9-parameter, dispersion-corrected GGA functional,[51] in combination with the def2-SVP split valence basis set.[52] The Q-Chem program package is used for all DFT calculations on H$_2$–Pc and Cu–Pc.[53] This higher level treatment of the electronic structure is compared to the results obtained with GFN-$x$TB, a semiempirical tight binding model conceived by the Grimme group,[54,55] and extended to other metal phatalocyanines (Cu–Pc, Ni–Pc, Co–Pc, and Fe–Pc) in the Supporting Information.

Rotational $g$-tensors are calculated via the density-fitted Hartree−Fock theory with gauge-including atomic orbitals (DF-HF-GIAO) as implemented in the Molpro package,[56−58] which also allows for computations of chemical shielding tensors and magnetizabilities.[59−61] To improve computational efficiency during tensor evaluations, basis set fitting is employed, and polarization functions are removed from Cu and H atoms.

An alternative treatment of magnetic moments caused by rotations and pseudorotations has been suggested by Ceseroli and Tosatti,[62] which is based on Berry−Phase calculations around closed orbits in the configuration space. For the sake of a direct comparison to previous studies on vibrationally induced molecular magnetism in smaller molecules, a selection of benzene and methane derivatives is discussed at the same level of theory.

## 3. RESULTS AND DISCUSSION

**3.1. Identification of IR-Active Pseudorotations.** In the first step, the minimum energy structures of six variants of phthalocyanine (H$_2$–Pc, Mn–Pc, Fe–Pc, Co–Pc, Ni–Pc, and Cu–Pc) as well as the benzene and methane derivatives are identified via DFT and a tight-binding ansatz as described above. The Hessians are evaluated at the same level of theory to obtain vibrational energies and their corresponding eigenmodes via matrix diagonalization. Structural information and detailed results are provided in the Supporting Information, which documents an acceptable performance of the GFN-$x$TB ansatz at a fraction of the computational effort of a DFT treatment. While vibrational energies deviate by less than 10% between the two approaches, larger deviations in terms of intensities are observed as they are known from more comprehensive studies.[63] However, correct identification of the relevant doubly degenerate vibrational modes is possible also at the GFN-xTB level of theory, making it a viable alternative for larger structures in the future.

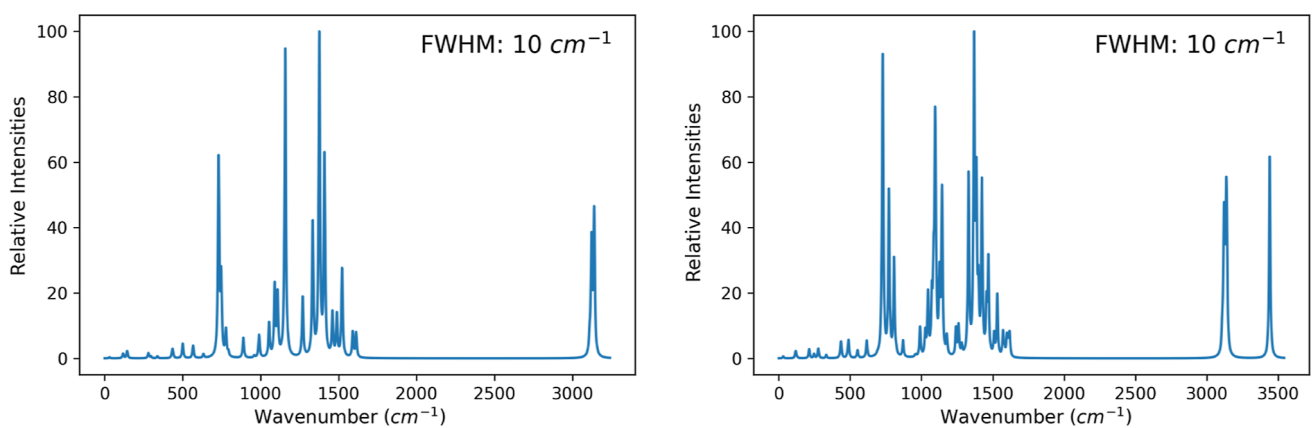

**Figure 2.** IR adsorption spectra of CuPc (left) and H$_2$Pc (right) as two representative phthalocyanines, obtained from DFT calculations. A Lorentzian line shape function with an fwhm of 10 cm$^{-1}$ was applied. Vibrations above 3000 cm$^{-1}$ correspond to motions of H atoms; the larger peaks around 1000 cm$^{-1}$ are mostly planar motions of N and C atoms.

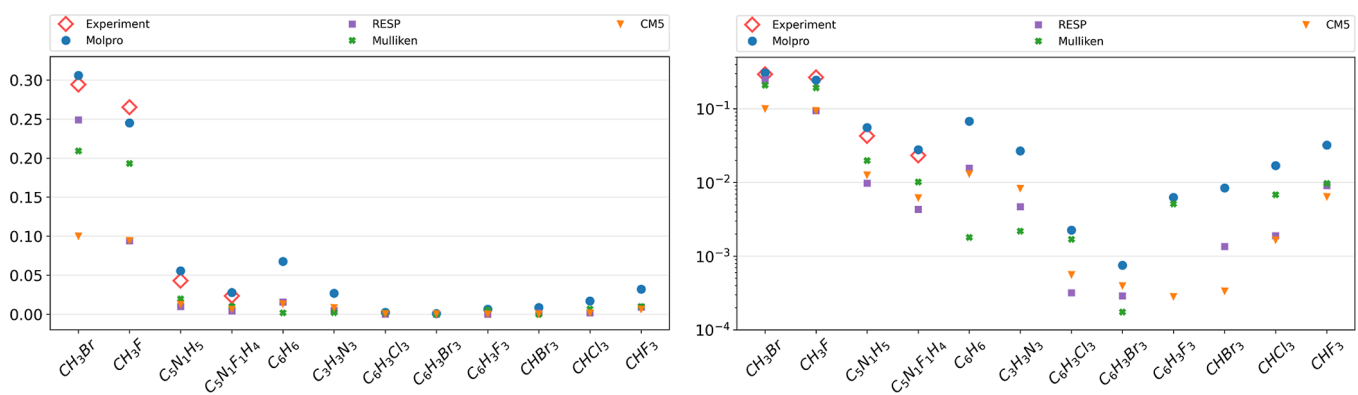

**Figure 3.** Absolute values of calculated and measured rotational g-factors, given in units of the nuclear magneton on linear (left) and logarithmic (right) scales. Theoretical values have been obtained with the Molpro via density-fitted Hartree−Fock theory with gauge-including atomic orbitals or by assuming a rigid rotation of various partial charges (RESP, CM5, and Mulliken) assigned to each atomic site.

For H$_2$Pc and CuPc, the corresponding calculated IR adsorption spectra are plotted in Figure 2 as a representative of the whole group. The obtained spectra are in excellent agreement with experimental data;[64] see the Supporting Information for a detailed comparison and spectral data of the other species.

In both cases, the spectrum can be broken into two parts, a lower part around 1000 cm$^{-1}$, which is mostly due to planar motion of the nitrogen or carbon atoms, and an upper part around 3100 cm$^{-1}$, which corresponds to planar motions of the hydrogen atoms of the benzene moieties. Qualitatively, but also in terms of line positions, the latter part of the spectrum is very closely related to the degenerate pairs of symmetric and antisymmetric C−H stretching vibrations of benzene (located at 3112 and 3125 cm$^{-1}$ at the same level of theory, respectively). The large peaks clearly correspond to doubly degenerate vibrations or pseudorotations, except for the peak near 700 cm$^{-1}$, which is the result of a nondegenerate out-of-plane motion of the N atoms. The additional peak at 3400 cm$^{-1}$ in the case of H$_2$Pc is unique in the series as it corresponds to the motion of the two H atoms at the center of the structure.

**3.2. Rotational g-Factors Evaluation and Benchmarking.** We start with a series of benchmark calculations for experimentally and theoretically studied benzene and methane derivatives. Our results are summarized in Figure 3. Results obtained with Molpro show an average deviation of less than 10% from the experimental values known for the three components of the rotational g tensor, which documents sufficient accuracy for the purpose of this article.[45,65,66]

In the next step, we evaluate the rotational g-factors of H$_2$Pc and CuPc for a rotation along the z-axis, i.e., perpendicular to the molecular plane, as they are a necessary ingredient for the evaluation of $g_{vib}$ according to the method of Moss and Perry. Interestingly, rounded to the accuracy mentioned above, identical values of $g_{rot}$ = 0.015 are obtained for both of them, which indicates a negligible variation over the whole series of metal atoms at this level of theory. On the other hand, a marginal impact of the central metal atom on the rotational g-factor is not fully unexpected, given the fact that it lies exactly on the rotational axis.

**3.3. Vibrational g-Factors.** After the calculation of rotational g-factors and the diagonalization of the molecular Hamiltonian, vibrational g-factors can be obtained either from eq 5, following the derivation of Moss and Perry, or via eq 10 with the help of fractional charges. While the choice of Born effective charges is straightforward from a solid state perspective, where a suitable description of effective macroscopic magnetic fields of bulk structures is desired, a correct treatment of intramolecular field geometries might necessitate a different approach. In the first step, we keep the concept of representative point charges but apply different charge localization strategies to the molecular benchmark set. We start with charges derived from the atomic polar tensor (APT), the molecular analogue to the Born effective charge tensor. As suggested by Milani et al.,[67] only the $z$-component of the APT is used. In addition, partial charges derived from the electron density[68] (Charge model 5, denoted as 'CM5' in the legend), from fits of the electrostatic potential[69] ("RESP"), as well as standard Mulliken charges are used as input in eq 10. Other partial charges, e.g., derived from the APT such as GAPT charges,[70] have been tested as well but provide inferior results and are therefore not discussed any further.

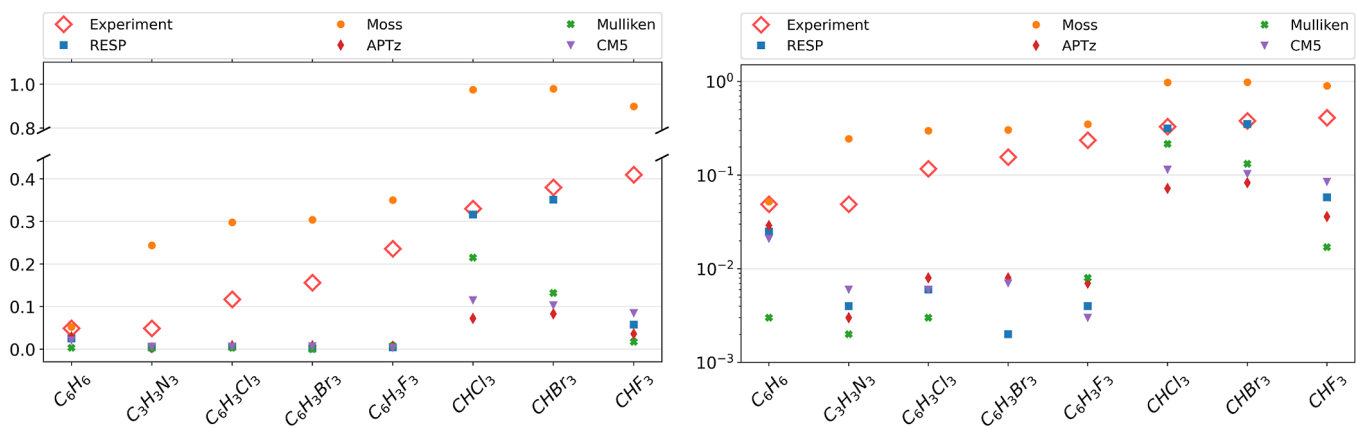

**Figure 4.** Absolute values of vibrational $g$-factors, obtained with all methods described in the text and compared to the experimentally confirmed eigenmodes at the following positions (from left to right, in cm$^{-1}$): 1480, 1545, 1565, 1550, 1610, 1213, 1141, and 1371. The figures show linear (left) and logarithmic (right) scales. Experimental results are extracted from refs 66 and, 71−73.

Our results are summarized in Figure 4. As becomes obvious from this comparison, all methods based on partial charges underestimate the vibrational $g$-factor substantially and are not even able to reproduce trends in the experimental data. A reason for this discrepancy may be an oversimplification that is intrinsic to all fractional charge models, namely, the assumption of point-like charge concentrations, a drastic simplification, even within a classical treatment, given the continuous character of the electron density distribution. For the sake of completeness, all models based on partial charges have been tested for their ability to predict rotational $g$-factors as well (see Figure 3), also without any success.

The ansatz of Moss and Perry, on the other hand, provides a fair reproduction of trends, although the results are still far away from quantitative agreement, in particular for the methane derivatives in the test set. Like all other methods mentioned above, their model also does not account for the substantial changes in the electron density that appear during molecular vibration (see later sections). However, a simple scaling by a factor of 0.404 reduces the relative root-mean-square deviation of $g_{vib}$ to 25%; see the Supporting Information for details.

After this comparison of theoretical models to known benchmark sets that offer, at least for some systems, also experimental verification, we switch back to the discussion of phthalocyanines. Our findings are summarized in Table 1, which contains energies, relative intensities, electric dipole moments, magnetic dipole moments, and vibrational $g$-factors, calculated with the method of Moss and Perry, for the most promising pseudorotations. Due to intrinsic numerical inaccuracies, some of the modes are not perfectly degenerate and show minimal deviations in the range of a few wavenumbers. In these cases, the average is listed in the table. Mode indices refer to the energetic order from the lowest to highest. Note that in the case of H$_2$Pc, vibrational degeneracies are only approximate due to the slightly reduced molecular symmetry, and both eigenmodes may have largely different intensities. Therefore, the $g$-factors are not calculated for H$_2$Pc. Entries for $\mu_{el}$, the electric dipole moment, have been evaluated at an amplitude of $\Delta Q = \sqrt{\frac{\hbar}{m_{red}\omega}\left(1 + \frac{1}{2}\right)}$, which corresponds to the standard deviation of a vibrational mode in its first excited state. The values of $\mu_{mag}$ have been calculated assuming an excitation to the first vibrationally excited state, i.e., $\langle G_t \rangle = \hbar$.

**Table 1. Selected Pseudorotations of the Cu Phthalocyanine Obtained from DFT Calculations: Index (Sorted by Energy, the Lowest to Highest), Vibrational Frequency, Relative Intensity, Electric Dipole Moment, Magnetic Dipole Moment, $g$-Factor, and Induced Magnetic Moment of the Doubly Degenerate Vibrations[a]**

| index | $\omega$/cm$^{-1}$ | rel. int. | $\mu_{el}$/Debye | $\mu_{mag}$/MHz/T | $g_{vib}$ |
|---|---|---|---|---|---|
| | | | C$_6$H$_6$ | | |
| 3, 4 | 605 | 0 | 0.000 | 1.56 | 0.205 |
| 21, 22 | 1467 | 5 | 0.069 | 0.39 | 0.052 |
| | | | CuPc | | |
| 79, 80 | 891 | 6 | 0.110 | 1.17 | 0.135 |
| 107, 108 | 1160 | 96 | 0.378 | 0.50 | 0.066 |
| 122, 123 | 1379 | 100 | 0.251 | 0.67 | 0.089 |
| 138, 139 | 1526 | 27 | 0.176 | 1.40 | 0.184 |
| 163, 164 | 3144 | 38 | 0.145 | 0.01 | 0.001 |

[a]For the magnetic moment, an excitation to the first excited state is assumed. Pseudorotations of benzene are listed as well for comparison.

The non-IR-active pseudorotation of benzene (experimentally found at 606 cm$^{-1}$, see ref 47), picked by Ceresoli and Tosatti,[62] is also included in the table for the sake of a direct comparison. In their work, a value of 0.79 is mentioned, evaluated via the Berry−Phase technique in a plane wave formalism using pseudopotentials. We note that this value, one of the very few estimates of $g_{\text{vib}}$ available in the literature, differs from our value calculated via eq 5. However, it is almost reproduced if a factor of $\sqrt{m_i}$ is added to the Coriolis coupling constant, which might be due to an incorrect normalization of the corresponding eigenmodes used in ref 62. Unfortunately, there is no established standard of eigenmode normalization in current computational chemistry codes since actual amplitudes are seldom of interest, and either type may be found in the standard output of common packages.

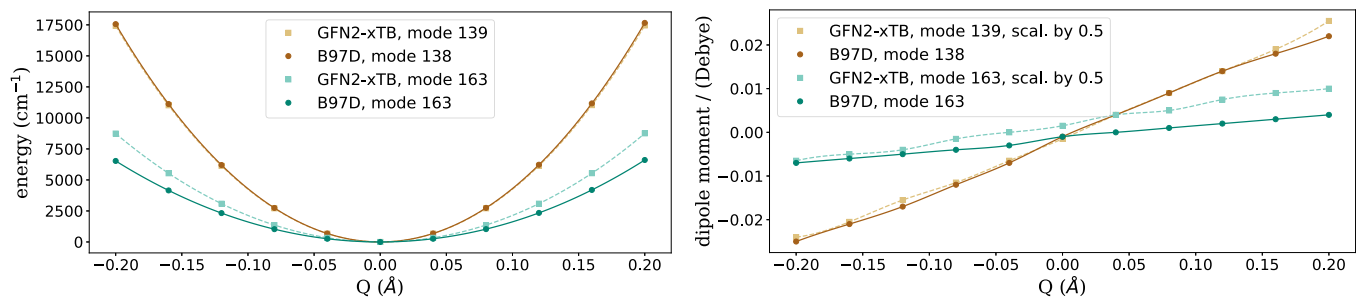

**Figure 5.** Scan over the energy (left) and the electric dipole moment (right) of CuPc as a function of the vibrational normal coordinate $Q$, corresponding to vibrational excitations at 1526 and 3144 cm$^{-1}$, calculated with DFT as well as the GFN2-$x$TB tight binding ansatz. The mode index mentioned in the legend refers to the energetic order after diagonalization of the corresponding Hessian matrix.

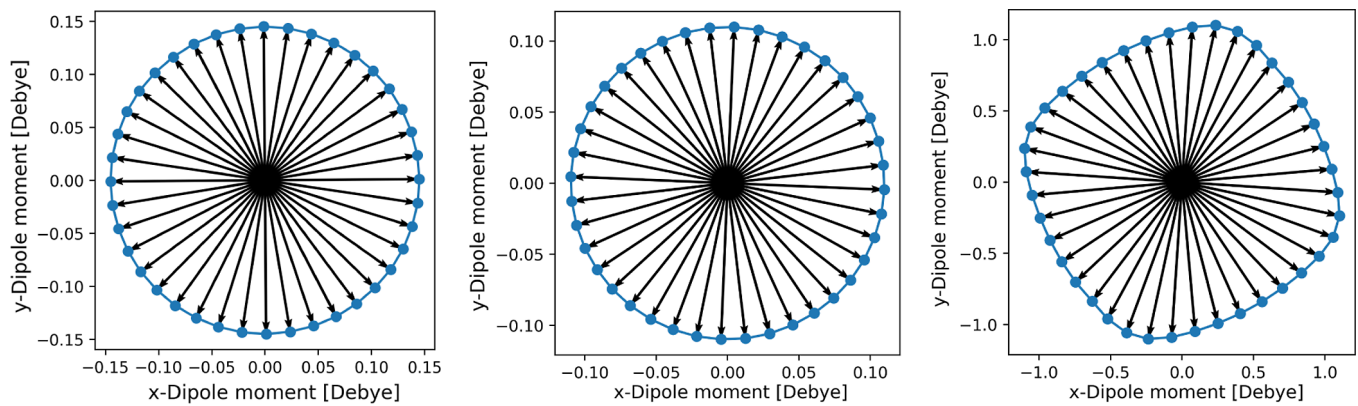

**Figure 6.** Plots of the electric dipole moment vector for pseudorotations at 3144 (left) and 891 cm$^{-1}$ (center), obtained by displacing all atoms according to each pair of degenerate, normalized eigenmodes $\vec{a}$, $\vec{b}$ via $\vec{p} = \Delta Q(\vec{a}\sin(\phi) + \vec{b}\cos(\phi))$ with $\vec{a}$ and $\vec{b}$ as the eigenvectors, in steps of $\delta\phi = \pi/20$. Vibrational amplitudes of $\Delta Q = 0.074$ Å and $\Delta Q = 0.077$ Å have been chosen, respectively, corresponding to the standard deviation of each first vibrationally excited state. For comparison, the pseudorotation at 891 cm$^{-1}$ is also presented at a much larger amplitude of $\Delta Q = 1$ Å (right) to indicate the onset of anharmonic behavior in the dipole moment.

**3.4. Electric Dipole Moment Dynamics.** Having investigated vibrational $g$-factors derived from either the vector coupling model of Moss and Perry or the more intuitive, yet qualitatively less applicable point charge models, we now look at vibrationally excited phthalocyanines from the perspective of electric dipole dynamics. For symmetry reasons, the molecule does not possess an electric dipole moment at the equilibrium geometry, but a distortion along a suitable, single vibrational eigenmode introduces a nonvanishing electric dipole moment. This can, e.g., be seen for the doubly degenerate vibrational excitations of CuPc at 1526 and 3144 cm$^{-1}$ shown in Figure 5, where energy and dipole moment predictions obtained with DFT and the GFN2-$x$TB tight-binding model are compared.

Both quantities are plotted as a function of one of the two corresponding (degenerate) normal modes. The degenerate mode at 1526 cm$^{-1}$ mostly corresponds to the displacement of N atoms and the other corresponds to the motion of H atoms, with reduced masses of around 12.4 and 1.1 amu, respectively. The latter explains the appearance of a wider parabola for the mode at higher energy. Note the remarkable agreement of the potential energy between both methods, which is almost perfect for the mode at 1526 cm$^{-1}$. Regarding the dipole moment, an almost linear behavior is found for both methods for small deviations from the equilibrium geometry. Interestingly, the dipole moment obtained with the GFN2-$x$TB method deviates almost exactly by a factor of 2 from the DFT result.

Combining two degenerate vibrational modes of CuPc with a phase shift of $\pi/2$, the linear behavior observed in Figure 5 must lead to a circular motion of the total molecular dipole moment, with its angular velocity determined by the frequency of the two degenerate eigenmodes; see the Supporting Information for further details. An example is displayed in Figure 6, which illustrates the degenerate modes of CuPc at 891 and 3144 cm$^{-1}$. However, note that this (for small deviations from the equilibrium geometry) almost perfectly circular motion of the molecular electric dipole moment does not necessarily translate into circular motions of individual atoms, as can be seen from Figure 7, where the degenerate normal modes are projected back onto the motion of each atom. These two examples of a pseudorotation in phthalocyanines are significantly different in character: One corresponds to a combined motion of all atoms and contains several circular trajectories, while the other corresponds mostly to C−H stretching in the benzene moieties and essentially linear nuclear trajectories.

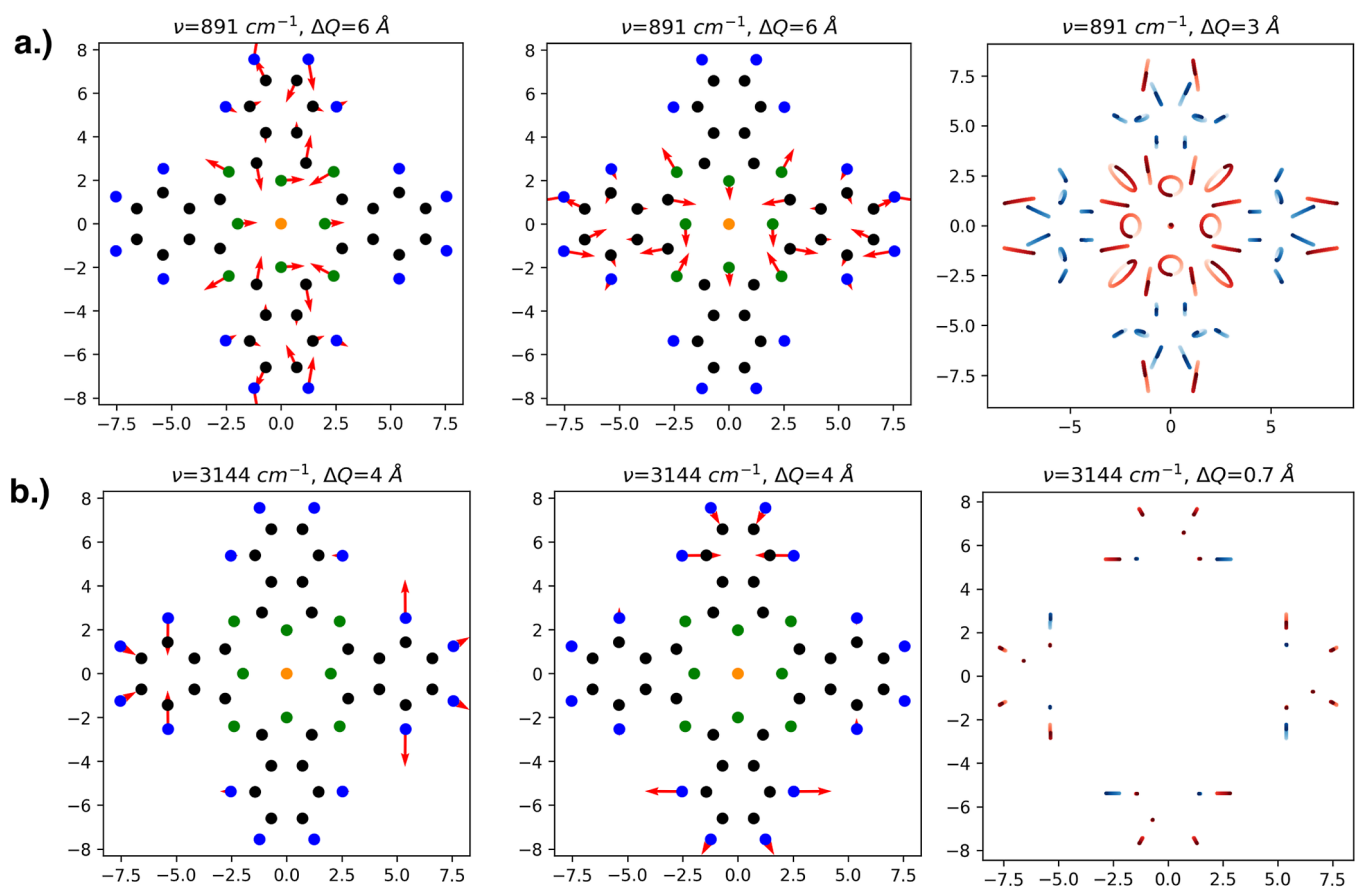

**Figure 7.** Pseudorotations of CuPc, realized by a linear combination of vibrational eigenmodes at (a) 891 and (b) 3144 cm$^{-1}$. The first and second columns show the two modes and the third column shows the corresponding pseudorotation. Displacements are indicated by red arrows; right and left circularly moving atoms are indicated by red and blue trajectories, respectively. Exaggerated vibrational amplitudes $\Delta Q$ were chosen for better illustration. The vibration at lower energy mostly consists of collective motions involving all atoms, while the selected pseudorotation at higher energy represents mostly C−H stretching in the benzene moieties. Qualitatively, this description is the same for all phthalocyanines under study.

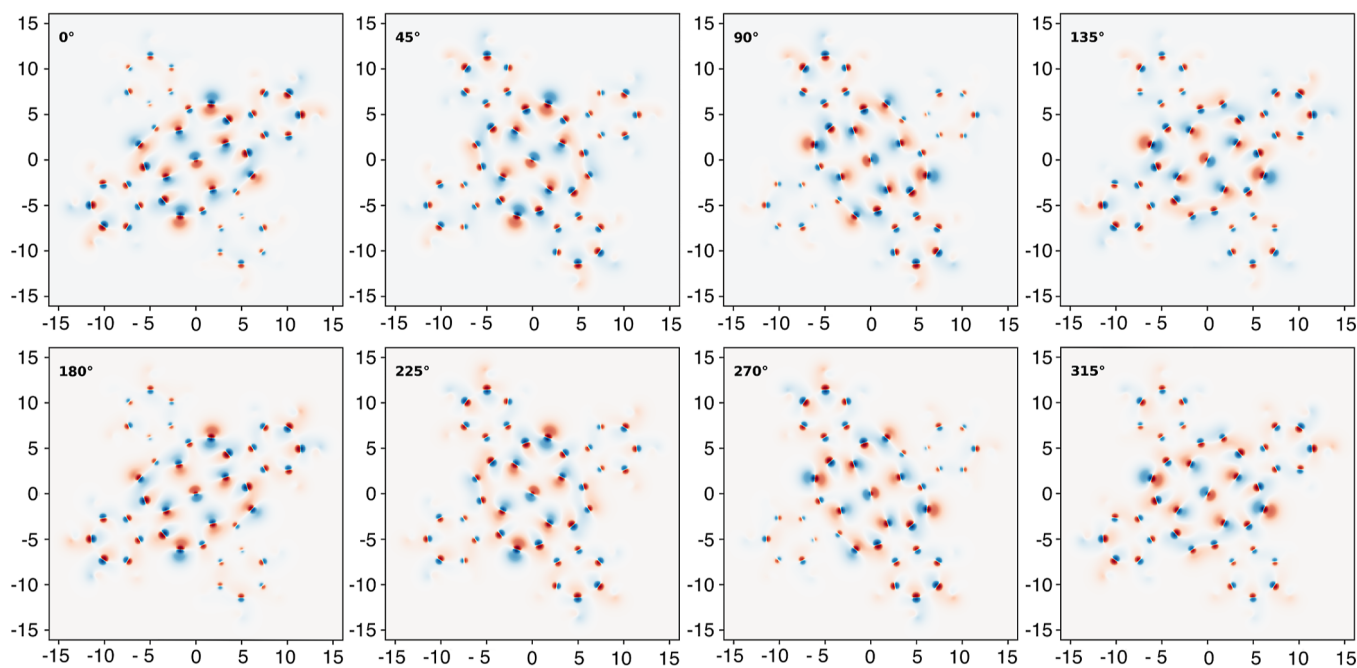

**Figure 8.** Changes in the electron density distribution of CuPc upon pseudorotation at 1526 cm$^{-1}$, in steps of 45 degrees and with amplitude $\Delta Q$ according to the standard deviation of the first excited state, made visible through a projection onto the molecular plane. Positions are given in Bohr; the electron density has been plotted logarithmically for a meaningful illustration of electron charge displacement.

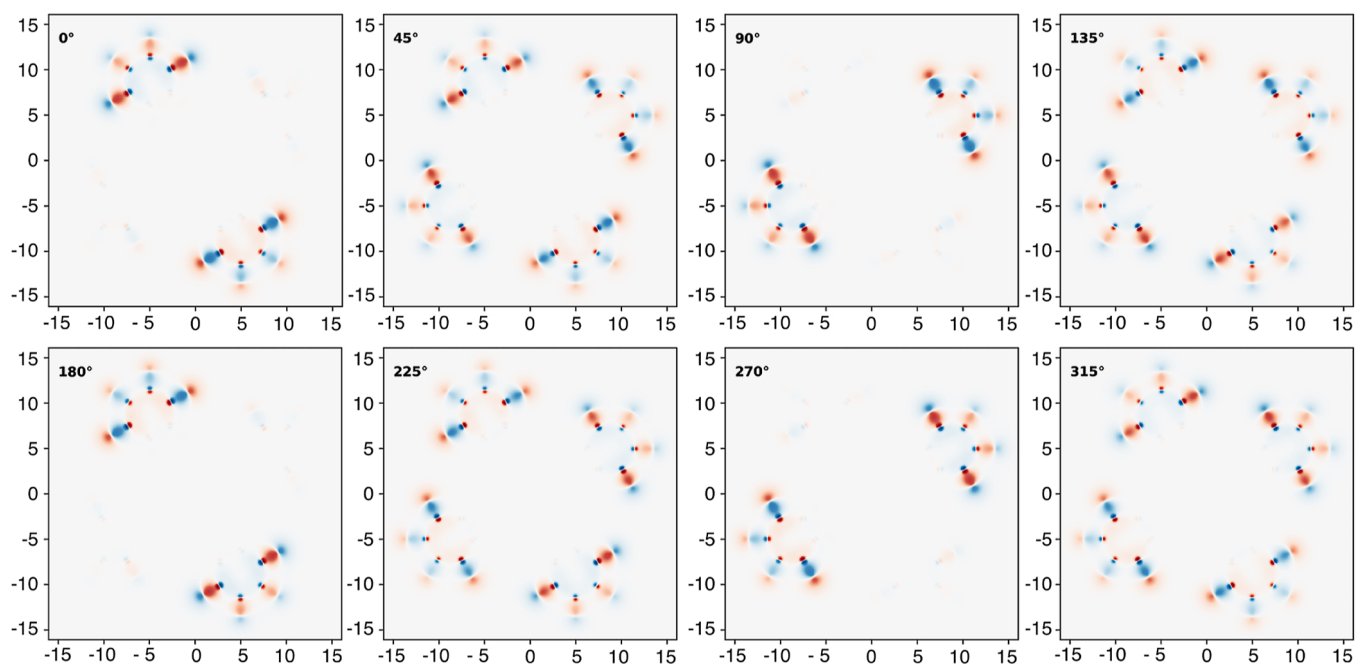

**Figure 9.** Changes in the electron density distribution of CuPc upon pseudorotation at 3144 cm$^{-1}$, in steps of 45 degrees and with amplitude $\Delta Q$ according to the standard deviation of the first excited state, made visible through a projection onto the molecular plane. Positions are given in Bohr; the electron density has been plotted logarithmically for a meaningful illustration of electron charge displacement.

It is equally problematic to state the reverse argument, namely, that the generation of a magnetic moment is hinged on a rotating electric dipole: For example, the pseudorotation mentioned above, selected by Ceresoli and Tosatti,[62] yields a nonzero vibrational g-factor, although it is based on two degenerate modes that are not IR-active. Introducing an evaluation method based on the Berry phase, they reveal a surprisingly low screening of the rotating ions by the electrons despite the aromaticity of the molecule, and a more complex behavior of the electric dipole moment is predicted. We will return to these earlier findings when discussing electron density fluctuations upon pseudorotation. Furthermore, even if an electric dipole moment exists and is only changing in orientation, the magnetic moment generated by the rotating centers of positive and negative charges will depend on their actual distance from the center of mass and the actual amount of fractional charge, none of which are unambiguously retrievable from the electric dipole moment alone. Plots of the motion of the two centers of charge during selected pseudororations are included in the Supporting Information, revealing that a dipole vector constructed from these two central moments is also performing a circular translation besides an actual rotation.

All of these findings point in the same direction: Local molecular magnetic fields, created through excitation of vibrational modes, are badly characterized by a single "vibrational g-factor" in the first place as the corresponding magnetic field geometry may be far too complex to be well described by a simple dipole term. As a consequence, vibrationally induced hyperfine splittings as well as chemical shifts, even those of identical nuclei, will differ with respect to the position of each nucleus within the same molecule. We note that these findings represent an analogy to atomic site-dependent g-factors as they are observable in the case of conventional Zeeman splitting, caused by an obviously inhomogeneous electron spin density of an open-shell molecule.[74] Interestingly, the latter had been proven in 2015 on Mn phthalocyanine, yet another member of the phthalocyanine family, but of quartet multiplicity.

**3.5. Electron Density Dynamics.** A last investigation of electric charge dynamics, this time with the highest possible resolution, brings us to the discussion of electron density fluctuations upon pseudorotation. For two degenerate modes, at 1526 and at 3144 cm$^{-1}$, changes of the electron density distribution upon vibrational excitation, evaluated with DFT using the B97D functional, are plotted in Figures 8 and 9. Again, a vibrational amplitude of $\Delta Q = \sqrt{\frac{\hbar}{m_{\text{red}}\omega}\left(1 + \frac{1}{2}\right)}$ has been chosen for display. An excess or lack of electron density in comparison to the unperturbed density at the equilibrium geometry is printed in blue or red, respectively. The electron density has been integrated in the z-direction, so the plots can be understood as projections of the electron density onto the molecular plane. In order to visualize the minute differences in the overall charge distribution, the density is plotted on a logarithmic scale. Otherwise, the substantial relocalization of atom-centered electron charge densities ("core" electrons) in the course of atomic displacements would have hampered the readability of the illustrations.

From Figure 8, it becomes immediately obvious that pseudorotations in macrocycles do cause periodic, but complicated patterns of charge displacements, which confirms to some extent the findings of ref 62 for benzene. In the case of the pseudorotation of CuPc at about 1526 cm$^{-1}$, where mostly C and N atoms are involved, the charge perturbations along the conjugated double bonds of the inner rings are dominating but far too complex to be approximated by a single rotating dipole. Instead, the electron density is constantly rotating along with the displaced atoms. The behavior of the electron density is particularly interesting in regions near the metal center: an almost perfectly circular motion of electron density is driven by the periodic vibration of the central atom. However, this motion of the central metal atom is present in only a few eigenmodes, as can be seen by comparing Figures 8 and 9. For the pseudorotation at about 3144 cm$^{-1}$, the electron density is modulated only at the hydrogen atoms, which are showing an almost perfectly linear type of motion (see also Figure 7). According either to Moss and Perry or to the "multiferroism" ansatz of Juraschek, linear motions do produce a vibrational $g$-factor of 0 since the cross product of the eigenvectors vanishes. As has been indicated already before, this is in harsh contrast to the oversimplified but intuitive interpretation of a (confirmed) rotating electric dipole moment as the generator of a molecule-wide magnetic dipole field. A definitive answer regarding the actual geometry of this vibrationally induced magnetic field can only be provided by experiments, which brings us to the next section, discussing possible observables for measurements in the laboratory.

### 3.6. Vibrationally Detuned Chemical Shielding.

If vibrationally induced magnetic fields, created via IR excitation of pairwise degenerate vibrational modes, are a physical reality, they should manifest themselves experimentally through comparably weak "vibrationally" induced splittings, e.g., in the hyperfine structure of the central metal atom. Of the centers investigated in this study, the stable and most abundant isotopes $^{55}_{25}$Mn, $^{59}_{27}$Co and $^{63}_{29}$Cu would be suitable candidates, in principle, for measurements of an optically induced, "vibrational" hyperfine splitting due to their nonzero nuclear spins of 5/2, 7/2, and 3/2.

However, the accurate description and measurement of hyperfine splittings in molecules are highly challenging. Instead, we propose a much simpler experimental approach based on standard NMR measurements of chemical shifts: a vibrationally induced magnetic field, activated through the photoexcitation of a suitable, IR-active pseudorotation, should provoke a significant change in the chemical shift observed for the resonance of each atomic nucleus simply due to the superposition of the external and intramolecular magnetic fields. These shifts will be easily distinguishable from the well-known, minimal effects of vibrationally induced chemical shifts[75] stemming directly from changes of the electron density distribution due to zero point vibrational energy or a certain Boltzmann occupation of vibrational levels at a given temperature. Contrary to the latter, effects due to the IR excitation of a suitable pseudorotation will be optically triggered and can cause additional chemical shifts in either direction, simply by changing the phase difference appearing in eq 12 from $-\pi/2$ to $\pi/2$.

In Table 2, we summarize the shifting effects caused by selected pseudorotations for all distinguishable nuclei in Cu phthalocyanine. Given the similarities in the vibrational modes as well as the vibrational $g$-factors found for other representatives, this table may serve as a first, crude reference for the whole class of metal phthalocyanines. The index corresponds to the distinguishable occurrences of each element within the molecule, ranked with respect to distance $d$ from the center; see Figure 10. For N, H, and Cu, these values represent the vibrational chemical shift (as C does not possess a nuclear spin) at this external field strength and are thus measurable quantities. They are calculated for a vibrational excitation of $\langle G_t \rangle = \hbar$ in the respective mode. Note that signs can be positive or negative, depending on the actual orientation of the pseudorotation, but the relative sign between shifts stays the same. For the degenerate vibrations at 3144 cm$^{-1}$, the induced chemical shift is (similar to the corresponding vibrational $g$-factor) approximately 2 orders of magnitude smaller than that for the degenerate vibrations at 1526 cm$^{-1}$.

Table 2. Estimates of the Magnetic Field Induced by a Stimulated Pseudorotation of CuPc at 1526 cm$^{-1}$, with $\langle G_t \rangle = \hbar$, Given for the $z$-Direction, Evaluated at Each Distinguishable Nucleus, and Expressed as Chemical Shifts, i.e., in ppm Relative to an External Field $B_0 = 1$ T; See Figure 10 for Positional Information

| index | element | $B_{vib}/B_0$ | $d$/Å |
|---|---|---|---|
| 1 | Cu | 5.3 | 0.0 |
| 2 | N | 11.9 | 2.0 |
| 3 | C | 10.8 | 3.0 |
| 4 | N | 9.5 | 3.4 |
| 5 | C | 5.8 | 4.3 |
| 6 | C | 0.7 | 5.6 |
| 7 | H | −0.3 | 5.9 |
| 8 | C | −0.5 | 6.6 |
| 9 | H | 1.0 | 7.7 |

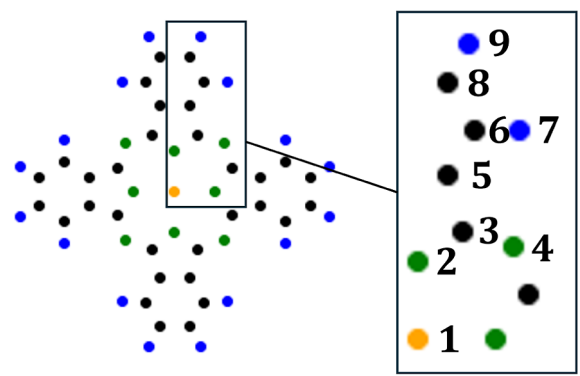

**Figure 10.** Positions of all distinguishable nuclei within CuPc, numbered according to their distance from the center; see Table 2 for the corresponding information on magnetic field strength.

## 4. CONCLUSIONS

Phthalocyanines, a highly versatile group of aromatic planar macrocyclic molecules, are well-known and appreciated for their magnetic properties. However, previous studies have been dedicated almost exclusively to spin Zeeman effects. In this study, the simultaneous, concerted excitation of two degenerate vibrational modes of a phthalocyanine molecule has been presented as a tool to create a localized, "molecular" magnetic field. By means of a computational study employing DFT, suitable vibrational excitations have been identified, and known theoretical descriptions have been presented, tested, and extended in order to allow for predictions of magnetic dipole moments and changes in the shielding tensors caused by these concerted vibrations.

Interestingly, although some IR-active pseudorotations do feature a rotating electric dipole moment, a direct conversion of this motion into a magnetic dipole moment, as suggested by classical electrodynamics, is not applicable: fluctuations of the electron density, observed during pseudorotation, suggest a much more complex geometry of the magnetic field even in these cases and indicate that magnetic field effects, e.g., on the nuclear spins of a molecule, must be site-specific due to substantial field inhomogeneities.

Fortunately, more intricate theoretical models have been known for decades, stemming from the golden era of NMR spectroscopy, where the motions of all nuclei upon pseudorotation are tracked and translated into separate contributions to the magnetic dipole field. Here, the description of Moss and Perry from 1972 stands out, but a closely related, more pragmatic ansatz has been pushed forward by the solid state community, termed as the 'dynamical multiferroic effect', where electron motion is accounted for through an effective nuclear charge derived from the polarization tensor. Inspired by this approach, we employ its molecular analogue, the atomic polarization tensor, and test several alternative partial charge models, e.g., derived from the electron density or the Coulomb potential. However, while the original ansatz, combined with modern ab initio evaluations of vibrational eigenmodes and rotational g-factors, could be improved by a simple scaling, methods based on point charges do not agree with the experiment at all. Hence, further work is needed in terms of method development, as well as the generation of accurate experimental data. Nevertheless, applying our best model to the class of phthalocyanines, we make first predictions for changes in the chemical shifts upon photoexcitation of the most promising IR-active pseudorotations. Experimental detection of the latter should be straightforward as these shifts will be triggered by the optical excitation of suitable modes and their sign will be dependent on the phase angle between the two oscillations.

Although vibrational couplings to spin magnetism have been known for decades in microwave spectroscopy, our work proposes a connection between an optically active and therefore experimentally accessible vibrational excitation of a molecule and its magnetic field generated at specific sites; the predicted changes in magnetic shielding constants, if confirmed by experiment, can be interpreted as first measurements of a vibrationally induced magnetic field with intramolecular resolution. The ability to switch or change the direction of this molecular magnetic field by optical setups, in combination with the potential of thin film depositions of phthalocyanines, makes this group of molecules an exciting target of future studies, adding an entirely new facet to this branch of magnetochemistry.

## ASSOCIATED CONTENT

### sı Supporting Information

The Supporting Information is available free of charge at https://pubs.acs.org/doi/10.1021/jacs.4c01915.

> Derivation of spin−vibration coupling parameters and rescaled g-factors, comparison to GFN2-xTB tight binding results, IR spectra for all phthalocyanines where experimental data are available, additional electron density plots, dipole moment scans along an IR-active normal mode, optimized geometries of all phthalocynanine structures in the xyz-format, and excerpts of Molpro output files containing information on vibrational modes (PDF)

## AUTHOR INFORMATION

**Corresponding Author**
  Andreas W. Hauser − *Institute of Experimental Physics, Graz University of Technology, A-8010 Graz, Austria;*
   orcid.org/0000-0001-6918-3106; Phone: +43 (316) 873-8157; Email: andreas.hauser@tugraz.at; Fax: +43 (316) 873-108152

**Authors**
  Raphael Wilhelmer − *Institute of Experimental Physics, Graz University of Technology, A-8010 Graz, Austria;*
   orcid.org/0009-0005-0175-4239
  Matthias Diez − *Institute of Experimental Physics, Graz University of Technology, A-8010 Graz, Austria;*
   orcid.org/0000-0001-6330-3082
  Johannes K. Krondorfer − *Institute of Experimental Physics, Graz University of Technology, A-8010 Graz, Austria;*
   orcid.org/0009-0009-5006-6319

Complete contact information is available at:
https://pubs.acs.org/10.1021/jacs.4c01915

**Notes**
The authors declare no competing financial interest.

## ■ ACKNOWLEDGMENTS

This research was funded in whole by the Austrian Science Fund (FWF) [10.55776/P36903]. For the purpose of open access, the author has applied a CC BY public copyright licence to any Author Accepted Manuscript version arising from this submission. Further support by NAWI Graz and the use of HPC resources provided by the IT services of Graz University of Technology (ZID) as well as the Vienna Scientific Cluster (VSC) is gratefully acknowledged.

## ■ REFERENCES

(1) Tagami, K.; Tsukada, M. Current-controlled magnetism in T-shape tape-porphyrin molecular bridges. *Curr. Appl. Phys.* **2003**, *3*, 439−444.
(2) Rai, D.; Hod, O.; Nitzan, A. Magnetic fields effects on the electronic conduction properties of molecular ring structures. *Phys. Rev. B* **2012**, *85*, 155440.
(3) Hod, O.; Rabani, E.; Baer, R. Magnetoresistance of Nanoscale Molecular Devices. *Acc. Chem. Res.* **2006**, *39*, 109−117.
(4) Coronado, E.; Epstein, A. J. Molecular spintronics and quantum computing. *J. Mater. Chem.* **2009**, *19*, 1670−1671.
(5) Kudisch, B.; Maiuri, M.; Moretti, L.; Oviedo, M. B.; Wang, L.; Oblinsky, D. G.; Prud'homme, R. K.; Wong, B. M.; McGill, S. A.; Scholes, G. D. Ring currents modulate optoelectronic properties of aromatic chromophores at 25 T. *Proc. Natl. Acad. Sci. U.S.A.* **2020**, *117*, 11289−11298.
(6) Naaman, R.; Waldeck, D. H. Chiral-Induced Spin Selectivity Effect. *J. Phys. Chem. Lett.* **2012**, *3*, 2178−2187.
(7) Banerjee-Ghosh, K.; Dor, O. B.; Tassinari, F.; Capua, E.; Yochelis, S.; Capua, A.; Yang, S.-H.; Parkin, S. S. P.; Sarkar, S.; Kronik, L.; Baczewski, L. T.; Naaman, R.; Paltiel, Y. Separation of enantiomers by their enantiospecific interaction with achiral magnetic substrates. *Science* **2018**, *360*, 1331−1334.
(8) Khachatryan, B.; Devir-Wolfman, A. H.; Tzabari, L.; Tessler, N.; Vardeny, Z. V.; Ehrenfreund, E. Magnetophotocurrent in Organic Bulk Heterojunction Photovoltaic Cells at Low Temperatures and High Magnetic Fields. *Phys. Rev. Appl.* **2016**, *5*, 044001.
(9) Weiss, L. R.; Bayliss, S. L.; Kraffert, F.; Thorley, K. J.; Anthony, J. E.; Bittl, R.; Friend, R. H.; Rao, A.; Greenham, N. C.; Behrends, J. Strongly exchange-coupled triplet pairs in an organic semiconductor. *Nat. Phys.* **2017**, *13*, 176−181.
(10) Bayliss, S. L.; Weiss, L. R.; Mitioglu, A.; Galkowski, K.; Yang, Z.; Yunusova, K.; Surrente, A.; Thorley, K. J.; Behrends, J.; Bittl, R.; et al. Site-selective measurement of coupled spin pairs in an organic semiconductor. *Proc. Natl. Acad. Sci. U.S.A.* **2018**, *115*, 5077−5082.
(11) Huynh, U. N. V.; Basel, T. P.; Ehrenfreund, E.; Vardeny, Z. V. Transient Magnetic Field Effect of Photoexcitations in Donor−

Acceptor Organic Semiconductors. *J. Phys. Chem. Lett.* **2018**, *9*, 4544−4549.

(12) Xu, H.; Wang, M.; Yu, Z.-G.; Wang, K.; Hu, B. Magnetic field effects on excited states, charge transport, and electrical polarization in organic semiconductors in spin and orbital regimes. *Adv. Phys.* **2019**, *68*, 49−121.

(13) Rodgers, C. T.; Hore, P. J. Chemical magnetoreception in birds: The radical pair mechanism. *Proc. Natl. Acad. Sci. U.S.A.* **2009**, *106*, 353−360.

(14) Park, J.; Koehler, F.; Varnavides, G.; Antonini, M.-J.; Anikeeva, P. Influence of Magnetic Fields on Electrochemical Reactions of Redox Cofactor Solutions. *Angew. Chem., Int. Ed.* **2021**, *60*, 18295−18302.

(15) Kavet, R.; Brain, J. Cryptochromes in Mammals and Birds: Clock or Magnetic Compass? *Physiology* **2021**, *36*, 183−194.

(16) Pople, J. Molecular orbital theory of aromatic ring currents. *Mol. Phys.* **1958**, *1*, 175−180.

(17) Lazzeretti, P. Ring currents. *Prog. Nucl. Magn. Reson. Spectrosc.* **2000**, *36*, 1−88.

(18) Peeks, M. D.; Claridge, T. D. W.; Anderson, H. L. Aromatic and antiaromatic ring currents in a molecular nanoring. *Nature* **2017**, *541*, 200−203.

(19) Valiev, R. R.; Benkyi, I.; Konyshev, Y. V.; Fliegl, H.; Sundholm, D. Computational Studies of Aromatic and Photophysical Properties of Expanded Porphyrins. *J. Phys. Chem. A* **2018**, *122*, 4756−4767.

(20) Peeks, M. D.; Gong, J. Q.; McLoughlin, K.; Kobatake, T.; Haver, R.; Herz, L. M.; Anderson, H. L. Aromaticity and Antiaromaticity in the Excited States of Porphyrin Nanorings. *J. Phys. Chem. Lett.* **2019**, *10*, 2017−2022.

(21) Wannere, C. S.; Schleyer, P. v. R. How Do Ring Currents Affect 1H NMR Chemical Shifts? *Org. Lett.* **2003**, *5*, 605−608.

(22) Schleyer, P. v. R.; Maerker, C.; Dransfeld, A.; Jiao, H.; van Eikema Hommes, N. J. R. Nucleus-Independent Chemical Shifts: A Simple and Efficient Aromaticity Probe. *J. Am. Chem. Soc.* **1996**, *118*, 6317−6318.

(23) Bartolomé, J.; Monton, C.; Schuller, I. K.. In *Molecular Magnets: Physics and Applications*; Bartolomé, J., Luis, F., Fernández, J. F., Eds.; Springer Berlin Heidelberg: Berlin, Heidelberg, 2014; pp 221−245.

(24) Melville, O. A.; Lessard, B. H.; Bender, T. P. Phthalocyanine-Based Organic Thin-Film Transistors: A Review of Recent Advances. *ACS Appl. Mater. Interfaces* **2015**, *7*, 13105−13118.

(25) Cranston, R. R.; Lessard, B. H. Metal phthalocyanines: thin-film formation, microstructure, and physical properties. *RSC Adv.* **2021**, *11*, 21716−21737.

(26) Rowland, H. Magnetic effect of electric convection. *Am. J. Sci.* **1878**, *s3-15*, 30−38.

(27) Bersuker, I. *The Jahn−Teller Effect*; Cambridge University Press, 2006.

(28) Domcke, W.; Yarkony, D. R.; Köppel, H. *Advanced Series in Physical Chemistry: Conical Intersections*; World Scientific, 2004; Vol. 15.

(29) Domcke, W.; Yarkony, D. R.; Köppel, H. *Advanced Series in Physical Chemistry: Conical Intersections*; World Scientific, 2011; Vol. 17.

(30) Berry, R. S. Correlation of Rates of Intramolecular Tunneling Processes, with Application to Some Group V Compounds. *J. Chem. Phys.* **1960**, *32*, 933−938.

(31) Barth, I.; Manz, J.; Sebald, P. Spinning a pseudorotating molecular top by means of a circularly polarized infrared laser pulse: Quantum simulations for 114CdH2. *Chem. Phys.* **2008**, *346*, 89−98. Accurate determination of molecular spectra and structure

(32) Barth, I.; Bressler, C.; Koseki, S.; Manz, J. Strong Nuclear Ring Currents and Magnetic Fields in Pseudorotating OsH4 Molecules Induced by Circularly Polarized Laser Pulses. *Chem.—Asian J.* **2012**, *7*, 1261−1295.

(33) Barth, I. Translational Effects on Electronic and Nuclear Ring Currents. *J. Phys. Chem. A* **2012**, *116*, 11283−11303.

(34) Kanno, M.; Kono, H.; Fujimura, Y. Control of π-Electron Rotation in Chiral Aromatic Molecules by Nonhelical Laser Pulses. *Angew. Chem., Int. Ed.* **2006**, *45*, 7995−7998.

(35) Kanno, M.; Hoki, K.; Kono, H.; Fujimura, Y. Quantum optimal control of electron ring currents in chiral aromatic molecules. *J. Chem. Phys.* **2007**, *127*, 204314.

(36) Mineo, H.; Fujimura, Y. Quantum control of coherent π-electron ring currents in polycyclic aromatic hydrocarbons. *J. Chem. Phys.* **2017**, *147*, 224301.

(37) Jia, D.; Manz, J.; Paulus, B.; Pohl, V.; Tremblay, J. C.; Yang, Y. Quantum control of electronic fluxes during adiabatic attosecond charge migration in degenerate superposition states of benzene. *Chem. Phys.* **2017**, *482*, 146−159. Electrons and nuclei in motion - correlation and dynamics in molecules (on the occasion of the 70th birthday of Lorenz S. Cederbaum)

(38) Juraschek, D. M.; Fechner, M.; Balatsky, A. V.; Spaldin, N. A. Dynamical multiferroicity. *Phys. Rev. Materials* **2017**, *1*, 014401.

(39) Juraschek, D. M.; Spaldin, N. A. Orbital magnetic moments of phonons. *Phys. Rev. Materials* **2019**, *3*, 064405.

(40) Juraschek, D. M.; Narang, P.; Spaldin, N. A. Phono-magnetic analogs to opto-magnetic effects. *Phys. Rev. Research* **2020**, *2*, 043035.

(41) Moss, R.; Perry, A. The vibrational Zeeman effect. *Mol. Phys.* **1973**, *25*, 1121−1134.

(42) Coudert, L. H.; Ernst, W. E.; Golonzka, O. Hyperfine coupling and pseudorotational motion interaction in Na3. *J. Chem. Phys.* **2002**, *117*, 7102−7116.

(43) Hauser, A. W.; Pototschnig, J. V.; Ernst, W. E. A classic case of Jahn−Teller effect theory revisited: Ab initio simulation of hyperfine coupling and pseudorotational tunneling in the 12E' state of Na3. *Chem. Phys.* **2015**, *460*, 2−13.

(44) Mead, C. A. The molecular Aharonov—Bohm effect in bound states. *Chem. Phys.* **1980**, *49*, 23−32.

(45) Flygare, W. H. Magnetic interactions in molecules and an analysis of molecular electronic charge distribution from magnetic parameters. *Chem. Rev.* **1974**, *74*, 653−687.

(46) Townes, C.; Schawlow, A. *Microwave Spectroscopy*; Dover Books on Physics; Dover Publications, 1975.

(47) Herzberg, G. *Molecular Spectra and Molecular Structure*; van Nostrand Reinhold, 1945; Vol. 2.

(48) Gauss, J.; Ruud, K.; Helgaker, T. Perturbation-dependent atomic orbitals for the calculation of spin-rotation constants and rotational g tensors. *J. Chem. Phys.* **1996**, *105*, 2804−2812.

(49) Darling, B. T.; Dennison, D. M. The Water Vapor Molecule. *Phys. Rev.* **1940**, *57*, 128−139.

(50) Watson, J. K. Simplification of the molecular vibration-rotation hamiltonian. *Mol. Phys.* **1968**, *15*, 479−490.

(51) Grimme, S.; Antony, J.; Ehrlich, S.; Krieg, H. A consistent and accurate ab initio parametrization of density functional dispersion correction (DFT-D) for the 94 elements H-Pu. *J. Chem. Phys.* **2010**, *132*, 154104.

(52) Weigend, F.; Ahlrichs, R. Balanced basis sets of split valence, triple zeta valence and quadruple zeta valence quality for H to Rn: Design and assessment of accuracy. *Phys. Chem. Chem. Phys.* **2005**, *7*, 3297.

(53) Epifanovsky, E.; Gilbert, A. T. B.; Feng, X.; Lee, J.; Mao, Y.; Mardirossian, N.; Pokhilko, P.; White, A. F.; Coons, M. P.; Dempwolff, A. L.; et al. Software for the frontiers of quantum chemistry: An overview of developments in the Q-Chem 5 package. *J. Chem. Phys.* **2021**, *155*, 084801.

(54) Bannwarth, C.; Ehlert, S.; Grimme, S. GFN2-xTB—An Accurate and Broadly Parametrized Self-Consistent Tight-Binding Quantum Chemical Method with Multipole Electrostatics and Density-Dependent Dispersion Contributions. *J. Chem. Theory Comput.* **2019**, *15*, 1652−1671.

(55) Bannwarth, C.; Caldeweyher, E.; Ehlert, S.; Hansen, A.; Pracht, P.; Seibert, J.; Spicher, S.; Grimme, S. Extended tight-binding quantum chemistry methods. *Wiley Interdiscip. Rev. Comput. Mol. Sci.* **2021**, *11*, No. e1493.

(56) Werner, H.-J.; Knowles, P. J.; Knizia, G.; Manby, F. R.; Schütz, M., et al. *MOLPRO*. Version 2022.2, 2022, Stuttgart, Germany. https://www.molpro.net.

(57) Werner, H.-J.; Knowles, P. J.; Knizia, G.; Manby, F. R.; Schütz, M. Molpro: a general-purpose quantum chemistry program package. *Wiley Interdiscip. Rev. Comput. Mol. Sci.* **2012**, *2*, 242−253.

(58) Werner, H.-J.; Knowles, P. J.; Manby, F. R.; Black, J. A.; Doll, K.; Heßelmann, A.; Kats, D.; Köhn, A.; Korona, T.; Kreplin, D. A.; et al. The Molpro quantum chemistry package. *J. Chem. Phys.* **2020**, *152*, 144107.

(59) Loibl, S.; Manby, F. R.; Schütz, M. Density fitted, local Hartree−Fock treatment of NMR chemical shifts using London atomic orbitals. *Mol. Phys.* **2010**, *108*, 477−485.

(60) Loibl, S.; Schütz, M. NMR shielding tensors for density fitted local second-order Møller-Plesset perturbation theory using gauge including atomic orbitals. *J. Chem. Phys.* **2012**, *137*, 084107.

(61) Loibl, S.; Schütz, M. Magnetizability and rotational g tensors for density fitted local second-order Møller-Plesset perturbation theory using gauge-including atomic orbitals. *J. Chem. Phys.* **2014**, *141*, 024108.

(62) Ceresoli, D.; Tosatti, E. Berry-Phase Calculation of Magnetic Screening and Rotational g Factor in Molecules and Solids. *Phys. Rev. Lett.* **2002**, *89*, 116402.

(63) Pracht, P.; Grant, D. F.; Grimme, S. Comprehensive Assessment of GFN Tight-Binding and Composite Density Functional Theory Methods for Calculating Gas-Phase Infrared Spectra. *J. Chem. Theory Comput.* **2020**, *16*, 7044−7060.

(64) Kobayashi, T.; Du, J.; Feng, W.; Yoshino, K.; Tretiak, S.; Saxena, A.; Bishop, A. R. Observation of breather excitons and soliton in a substituted polythiophene with a degenerate ground state. *Phys. Rev. B* **2010**, *81*, 075205.

(65) Stolze, W. H.; Stolze, M.; Hübner, D.; Sutter, D. H. The Rotational Zeeman Effect in Fluorobenzene and the Molecular Quadrupole Moment in Benzene. *Z. Naturforsch. A* **1982**, *37*, 1165−1175.

(66) Flygare, W.; Benson, R. The molecular Zeeman effect in diamagnetic molecules and the determination of molecular magnetic moments (g values), magnetic susceptibilities, and molecular quadrupole moments. *Mol. Phys.* **1971**, *20*, 225−250.

(67) Milani, A.; Castiglioni, C. Atomic charges from atomic polar tensors: A comparison of methods. *J. Mol. Struct.* **2010**, *955*, 158−164.

(68) Marenich, A. V.; Jerome, S. V.; Cramer, C. J.; Truhlar, D. G. Charge Model 5: An Extension of Hirshfeld Population Analysis for the Accurate Description of Molecular Interactions in Gaseous and Condensed Phases. *J. Chem. Theory Comput.* **2012**, *8*, 527−541.

(69) Wang, J.; Cieplak, P.; Kollman, P. A. How well does a restrained electrostatic potential (RESP) model perform in calculating conformational energies of organic and biological molecules? *J. Comput. Chem.* **2000**, *21*, 1049−1074.

(70) Richter, W. E.; Duarte, L. J.; Bruns, R. E. Are "GAPT Charges" Really Just Charges? *J. Chem. Inf. Model.* **2021**, *61*, 3881−3890.

(71) Wang, B.; Tam, C. N.; Keiderling, T. A. Vibrational Zeeman effect for the $\nu_4$ mode of haloforms ($HCX_3$) determined by magnetic vibrational circular dichroism. *Phys. Rev. Lett.* **1993**, *71*, 979−982.

(72) Wang, B.; Keiderling, T. A. Measurement of the vibrational Zeeman effect for HCF3 using magnetic vibrational circular dichroism. *J. Chem. Phys.* **1994**, *101*, 905−911.

(73) Devine, T. R.; Keiderling, T. A. Magnetic vibrational circular dichroism of benzene and 1,3,5-substituted derivatives. *J. Phys. Chem.* **1984**, *88*, 390−394.

(74) Liu, L.; Yang, K.; Jiang, Y.; Song, B.; Xiao, W.; Song, S.; Du, S.; Ouyang, M.; Hofer, W. A.; Castro Neto, A. H.; Gao, H.-J. Revealing the Atomic Site-Dependent g Factor within a Single Magnetic Molecule via the Extended Kondo Effect. *Phys. Rev. Lett.* **2015**, *114*, 126601.

(75) Sabzyan, H.; Buzari, B. Theoretical study of the contribution of vibrational motions to nuclear shielding constants. *Chem. Phys.* **2008**, *352*, 297−305.

## NOTE ADDED AFTER ASAP PUBLICATION

This paper was originally published ASAP on May 14, 2024, with an error in the Figure 4 caption. The corrected version was reposted on May 17, 2024.

# Supporting Information:

# Molecular pseudorotation in phthalocyanines as a tool for magnetic field control at the nanoscale

Raphael Wilhelmer, Matthias Diez, Johannes K. Krondorfer and Andreas W. Hauser*

*Institute of Experimental Physics, Graz University of Technology, Petersgasse 16, A-8010 Graz, Austria. Phone: +43 (316) 873-8157, Fax: +43 (316) 873-108152*

E-mail: andreas.w.hauser@gmail.com

In the first section of this Supporting Information, we provide the details on the calculation of vibrational g-factors and spin-vibration coupling using perturbation theory and the rescaled vibrational g-factors. Section S2 gives a summary of vibrational frequencies in tabulated form, obtained with the GFN-xTB tight binding ansatz of the Grimme group.[1,2] In Section S3, IR spectra are provided for all phthalocyanines, calculated with DFT via the Q-Chem program package,[3] and compared to the corresponding GFN-xTB results. Additional data, comparing the calculated to experimental values, is also presented. This is followed by a series of graphical illustrations of electron density changes in Cu phthalocyanine upon selected pseudorotations, evaluated with Q-Chem. In Section S5, dipole moment scans along a IR-active normal mode of $H_2Pc$ are presented for a comparison between the two computational methods, i.e. the tight binding ansatz and the results obtained with dispersion-corrected DFT using the PBE functional. Additionally, the origin of the rotating dipole moment is illustrated via the center of charges and different partial charges are investigated in terms of their ability to reproduce the electric dipole moment. Finally, in Section S6, optimized geometries of all phthalocynanine structures are listed in xyz-format, followed by excerpts of Q-Chem output files containing information on vibrational modes of CuPc evaluated with DFT in Section S7.

# Contents

# S1 Derivation of vibrational coupling parameters and rescaled g-factors

## S1.1 Derivation of vibrational g-factors and spin-vibration coupling parameters

Considering the effect of nuclear rotation and vibration, a complete description of nuclear motion is possible via center of mass coordinates, Euler angles and vibrational coordinates evaluated at the minimum of a selected potential energy surface. After expressing the classical kinetic energy in these coordinates, one obtains a metric, which may be used to construct a corresponding quantum Hamiltonian for the entire molecular system by canonical quantization.[4,5] This procedure yields the total Hamiltonian

$$H = \sum_{i=0}^{N_e} \frac{\mathbf{p}_i^2}{2m_e} + V_c$$
$$+ \frac{1}{2}(\mathbf{J} - \mathbf{G} - \mathbf{L}_e)\Theta_{\text{eff}}^{-1}(\mathbf{J} - \mathbf{G} - \mathbf{L}_e) \quad \text{(S1)}$$
$$+ \frac{1}{2}\sum_r^{3N_N-6} P_r^2 - \frac{\hbar^2}{8}\sum_{\alpha=1}^{N_n} \Theta_{\text{eff},\alpha\alpha}^{-1},$$

where the first line corresponds to the electronic kinetic energy and all types of intramolecular Coulomb interactions, the second line corresponds to the rotational kinetic energy of the nuclear system in the respective coordinate frame, and the last line contains vibrational kinetic energy of the nuclei and a mass correction term, which is negligible. For small displacements from the equilibrium $\Theta_{\text{eff}}$ can be approximated by the nuclear inertia tensor $\Theta$ in the equilibrium configuration.

The nuclear rotational energy is relevant for vibrational and rotational coupling. Here $\mathbf{J} = \mathbf{L}_e + \mathbf{L}_N + \mathbf{G}$ denotes the total angular momentum, with electronic contribution $\mathbf{L}_e = \sum_{i=0}^{N_e} \mathbf{L}_i$ and nuclear contribution $\mathbf{L}_N$. The vibrational angular momentum $\mathbf{G}$ is only non-zero for degenerated vibrational excitation and can be written as

$$\mathbf{G} = \sum_\alpha \mathbf{G}_\alpha = \sum_{\alpha,t} \boldsymbol{\zeta}_{\alpha,t}(Q_{t_1}P_{t_2} - Q_{t_2}P_{t_1}) = \sum_t \boldsymbol{\zeta}_t G_t, \quad (S2)$$

with $t$ indexing the doubly degenerate vibrational excitations and $Q_{t_1}$, $Q_{t_2}$ denoting the respective normal coordinates. Furthermore, the scalar vibrational angular momentum $G_t$ for each degenerate pair of eigenmodes has been introduced. $\boldsymbol{\zeta}_t$ denotes the so-called Coriolis coupling constant and $\boldsymbol{\zeta}_{\alpha,t} = \boldsymbol{d}_{\alpha,t_1} \times \boldsymbol{d}_{\alpha,t_2}$ the respective contribution of nucleus $\alpha$, where $\boldsymbol{d}_{\alpha,t_i}$ denotes the normalized mass weighted displacement vector for vibrational mode $i$.

We can perform first order state correction with perturbation $-(\mathbf{J} - \mathbf{G})\Theta^{-1}\mathbf{L}_e$ for the electronic system to obtain

$$|0\rangle^1 = |0\rangle - \sum_{n \neq 0} \langle \mathbf{J} - \mathbf{G} \rangle \Theta^{-1} \frac{\langle n | \mathbf{L}_e | 0 \rangle}{E_0 - E_n} |n\rangle. \quad (S3)$$

The expression $\langle \mathbf{J} - \mathbf{G} \rangle$ denotes the expectation value with respect to the nuclear system. Usually, however, this term is treated classically. The electronic contribution to rotational and vibrational coupling parameters can now be calculated by evaluating the expectation value of a suitable interaction Hamiltonian in the perturbed state.

In order to obtain rotational and vibrational g-factors we consider the paramagnetic interaction Hamiltonian with an external field

$$H_B^{\text{para}} = \left( \frac{e}{2m_e}\mathbf{L}_e - \sum_\alpha \frac{Z_\alpha e}{2M_\alpha}\mathbf{L}_\alpha \right) \cdot \mathbf{B} \quad (S4)$$

. The angular momentum of nucleus $\alpha$ can be expressed as

$$\mathbf{L}_{\alpha,\text{com}} = \Theta^{(\alpha,\text{com})}\Theta^{-1}\left(\mathbf{J} - \mathbf{L}_e - \mathbf{G}\right) + \mathbf{G}_\alpha, \quad (S5)$$

with $\Theta^{(\alpha,\text{com})} = M_\alpha[(\mathbf{R}_\alpha - \mathbf{R}_{\text{com}})^2 \mathbb{1} - (\mathbf{R}_\alpha - \mathbf{R}_{\text{com}})(\mathbf{R}_\alpha - \mathbf{R}_{\text{com}})^T]$ the inertia tensor of nucleus $\alpha$ with respect to the center of mass. Using this yields the electronic and nuclear contribution to the rotationally and vibrationally induced magnetic moment are

$$\mu_{e,i} = -\frac{e}{2m_e} \left[ \langle 0| L_{e,i} |0\rangle - \sum_{k,l}(\Theta^{-1})_{kl}\langle J_k - G_k\rangle \sum_{n\neq 0} \frac{\langle 0| L_{e,i} |n\rangle \langle n| L_{e,l} |0\rangle + c.c.}{E_0 - E_n} \right] \quad (S6)$$

$$\boldsymbol{\mu}_N = e\sum_\alpha \frac{Z_\alpha}{2M_\alpha}\Theta^{(\alpha,\text{com})}\Theta^{-1}(\mathbf{J} - \mathbf{L}_e - \mathbf{G}) + \sum_t \sum_\alpha \frac{eZ_\alpha \boldsymbol{\zeta}_{\boldsymbol{\alpha},t}}{2M_\alpha} G_t. \quad (S7)$$

For a $\Sigma$ ground state, this can be written as $\boldsymbol{\mu}^{\text{ind}} = g^{\text{rot}}\mathbf{J} - \sum_t \boldsymbol{g}_t^{\text{vib}} G_t$, with the rotational and vibrational g-factors

$$\begin{aligned} g^{\text{rot}} &= -\frac{2m_e}{e}\xi^{\text{para}}\Theta^{-1} + e\sum_\alpha \frac{Z_\alpha}{2M_\alpha}\Theta^{(\alpha,\text{com})}\Theta^{-1} \\ \boldsymbol{g}_t^{\text{vib}} &= \sum_k (g^{rot}\boldsymbol{\zeta}_t) - \sum_\alpha \frac{eZ_\alpha(\boldsymbol{\zeta}_{\alpha,t})}{2M_\alpha}. \end{aligned} \quad (S8)$$

with $\xi^{\text{para}}$ the paramagnetic magnetizibility given by

$$\xi_{ij}^{\text{para}} = -\left(\frac{e}{2m}\right)^2 \sum_{n\neq 0} \frac{1}{E_0 - E_n}(\langle 0| L_{e,j} |n\rangle \langle n| L_{e,i} |0\rangle + c.c.) \quad (S9)$$

Equation S8 reproduces the formulas of vibrational and rotational g-factors from literature.[6,7]

The magnetic moment, however, does not contain information on the magnetic field distribution within the molecule. To obtain measurable quantities we calculate the rotationally and vibrationally induced NMR splitting for each nucleus. The corresponding interaction Hamiltonian

$$H_{\text{SO}} = \sum_\alpha \frac{\mu_0}{4\pi} e \left[ -\sum_i \frac{1}{m_e} \frac{\mathbf{L}_{i,\alpha}}{|\mathbf{R}_\alpha - \mathbf{r}_i|^3} + \sum_{\beta\neq\alpha} \frac{Z_\beta}{M_\beta} \frac{\mathbf{L}_{\beta,\alpha}}{|\mathbf{R}_\alpha - \mathbf{R}_\beta|^3} \right] \cdot \gamma_\alpha \mathbf{I}_\alpha \quad (S10)$$

is the spin-orbit coupling Hamiltonian with $\mathbf{L}_{i,\alpha}$ the electron angular momentum with respect to nucleus $\alpha$, $\mathbf{L}_{\beta,\alpha}$ the nuclear angular momentum of nucleus $\beta$ with respect to nucleus $\alpha$, the nuclear gyromagnetic ratio $\gamma_\alpha$ and the nuclear spin operator $\mathbf{I}_\alpha$. Analogously to the calculation of the g-factor we calculate the expectation value in the perturbed state and use a similar expression for $\mathbf{L}_{\beta,\alpha}$ as in Equation S5. In a $\Sigma$ ground state the rotationally and vibrationally induced magnetic field at nucleus $\alpha$ can then be written as

$$\mathbf{B}^{\text{ind}}_{\text{rot},\alpha} = \frac{2m_e}{e}\sigma^{\text{para}}_\alpha \Theta^{-1}\mathbf{J} + \frac{\mu_0 e}{4\pi}\sum_\beta \frac{Z_\beta}{M_\beta}\frac{\Theta^{(\beta,\alpha)}\Theta^{-1}\mathbf{J}}{|\mathbf{R}_\alpha - \mathbf{R}_\beta|^3}$$
$$\mathbf{B}^{\text{ind}}_{\text{vib},\alpha} = -\frac{2m_e}{e}\sigma^{\text{para}}_\alpha \Theta^{-1}\mathbf{G} + \frac{\mu_0 e}{4\pi}\sum_\beta \frac{Z_\beta}{M_\beta}\frac{\mathbf{G}_\beta - \Theta^{(\beta,\alpha)}\Theta^{-1}\mathbf{G}}{|\mathbf{R}_\alpha - \mathbf{R}_\beta|^3}$$

(S11)

with $\Theta^{(\beta,\alpha)}$ the inertia tensor of nucleus $\beta$ with respect to nucleus $\alpha$ and $\sigma^{\text{para}}_\alpha$ the paramagnetic shielding tensor at nucleus $\alpha$ given by

$$(\sigma^{\text{para}}_\alpha)_{kj} = \frac{\mu_0}{4\pi}\frac{e^2}{2m_e^2}\sum_i\sum_n \frac{1}{E_0 - E_n}\left[\langle 0|\frac{(L_{i,\alpha})_k}{|\mathbf{R}_\alpha - \mathbf{r}_i|^3}|n\rangle\langle n|(L_{e,\alpha})_j|0\rangle + c.c.\right] \quad \text{(S12)}$$

For the rotationally induced magnetic field, this is consistent with Ref. 7 but generalizes the result for pseudorotational excitations.

## S1.2 Rescaled vibrational g-factors

As explained in the main article, vibrational g-factors calculated by the method of Moss and Perry generally overestimate the vibrational g-factors. However, re-scaling the g-factors by a factor of 0.404 (which minimized the mean squared error of the g-factors to 0.05) leads to much better results. This can be seen in Figure S1, where the rescaled g-factors are plotted.

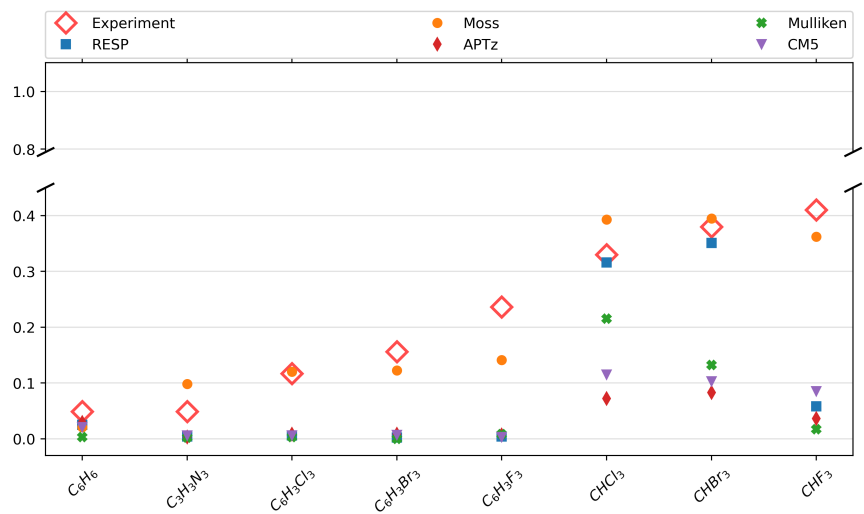

Figure S1: Vibrational g-factors calculated by the method of Moss and Perry as described in the main article and rescaled by a factor of 0.404.

# S2 GFN2-xTB data for comparison

This sections provides a quick overview of the vibrational frequencies obtained with the GFN2-xTB method.

Table S1: Strongly IR-active pseudorotations of benzene, $H_2$-, Mn-, and Fe-phthalocyanines obtained from GFN2-xTB calculations: index, vibrational frequency, relative intensity.

| index | $\omega$ / cm$^{-1}$ | rel. int. |
|---|---|---|
| \multicolumn{3}{c}{$C_6H_6$} | | |
| 3, 4 | 580 | 0 |
| 28, 29 | 3084 | 120 |
| \multicolumn{3}{c}{$H_2Pc$} | | |
| 121, 122 | 1319, 1339 | 219, 216 |
| 137, 138 | 1514, 1516 | 75, 228 |
| 155, 157 | 3085, 3091 | 253, 229 |
| 162, 166 | 3103, 3109 | 400, 205 |
| \multicolumn{3}{c}{MnPc} | | |
| 103, 104 | 1143 | 116 |
| 123, 124 | 1345 | 187 |
| 138, 139 | 1520 | 239 |
| 155, 156 | 3089 | 127 |
| 163, 164 | 3108 | 144 |
| \multicolumn{3}{c}{FePc} | | |
| 103, 104 | 1149 | 178 |
| 122, 123 | 1345 | 207 |
| 138, 140 | 1525 | 247 |
| 155, 156 | 3088 | 127 |
| 163, 164 | 3108 | 122 |

Table S2: Strongly IR-active pseudorotations of Co-, Ni-, and Cu-phtalocyanine obtained from GFN2-xTB calculations: index, vibrational frequency, relative intensity.

| index | $\omega$ / cm$^{-1}$ | rel. int. |
|---|---|---|
| CoPc | | |
| 103, 104 | 1149 | 130 |
| 122, 123 | 1348 | 154 |
| 138, 139 | 1534 | 231 |
| 155, 156 | 3089 | 269 |
| 163, 164 | 3108 | 283 |
| NiPc | | |
| 103, 104 | 1149 | 135 |
| 122, 123 | 1347 | 170 |
| 138, 139 | 1524 | 178 |
| 155, 156 | 3089 | 226 |
| 163, 164 | 3109 | 200 |
| CuPc | | |
| 102, 103 | 1156 | 107 |
| 123, 124 | 1345 | 188 |
| 139, 140 | 1515 | 163 |
| 155, 156 | 3088 | 300 |
| 163, 164 | 3107 | 262 |

# S3 IR spectra details

In this section we present and compare calculated spectra. Spectra calculated with Q-Chem employ the B97D GGA functional and the def2-SVP basis set. Spectra calculated using the GFN2-xTB approach are compared to the DFT results. Also, a comparison of the CuPc and $H_2$Pc DFT calculations to experiment[8–10] is provided.

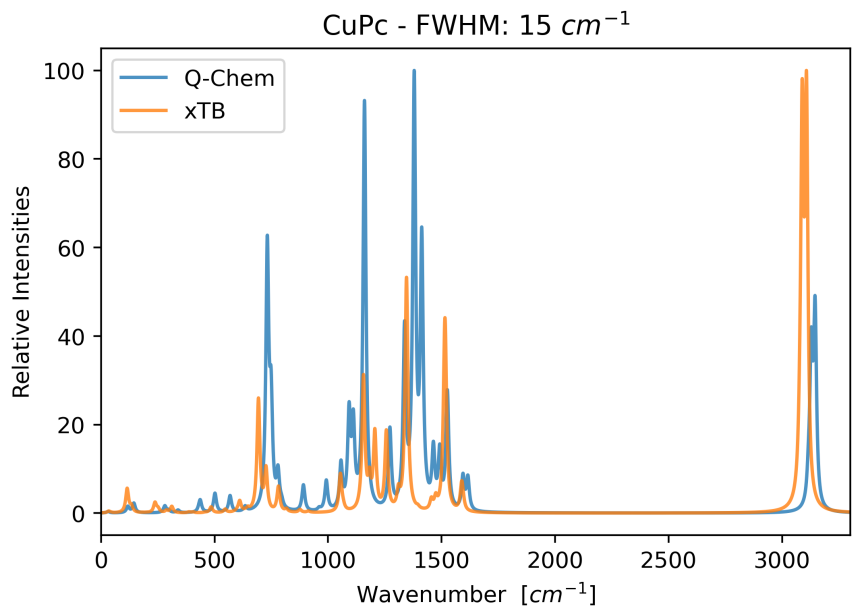

Figure S2: IR adsorption spectra of CuPc obtained from GFN-xTB and Q-Chem calculations (see main article). Vibrations above 3000 cm$^{-1}$ correspond to motions of H atoms; the larger peaks around 1000 cm$^{-1}$ are mostly planar motions of N atoms. A Lorenz line shape function with FWHM of 10 cm$^{-1}$ was applied.

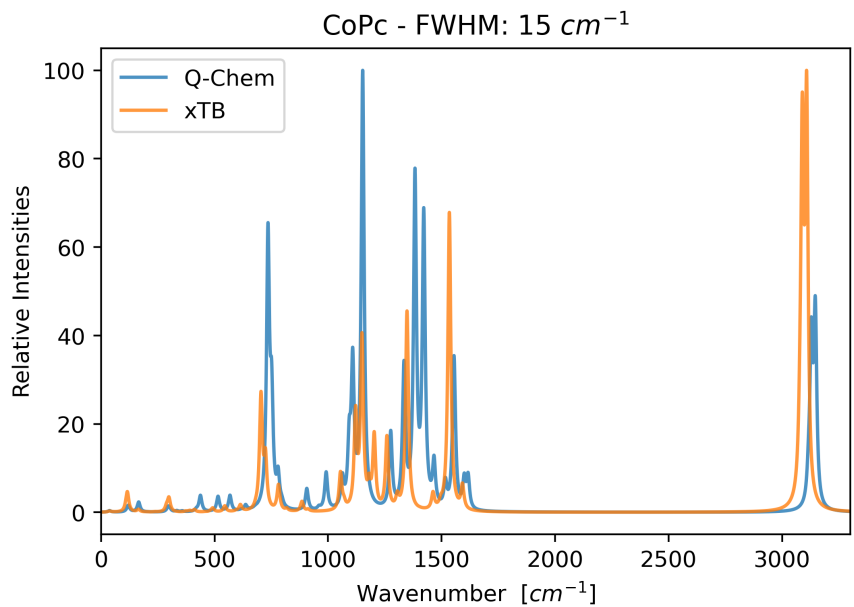

Figure S3: IR adsorption spectra of CoPc obtained from GFN-xTB and Q-Chem calculations (see main article). Vibrations above 3000 cm$^{-1}$ correspond to motions of H atoms; the larger peaks around 1000 cm$^{-1}$ are mostly planar motions of N atoms. A Lorenz line shape function with FWHM of 10 cm$^{-1}$ was applied.

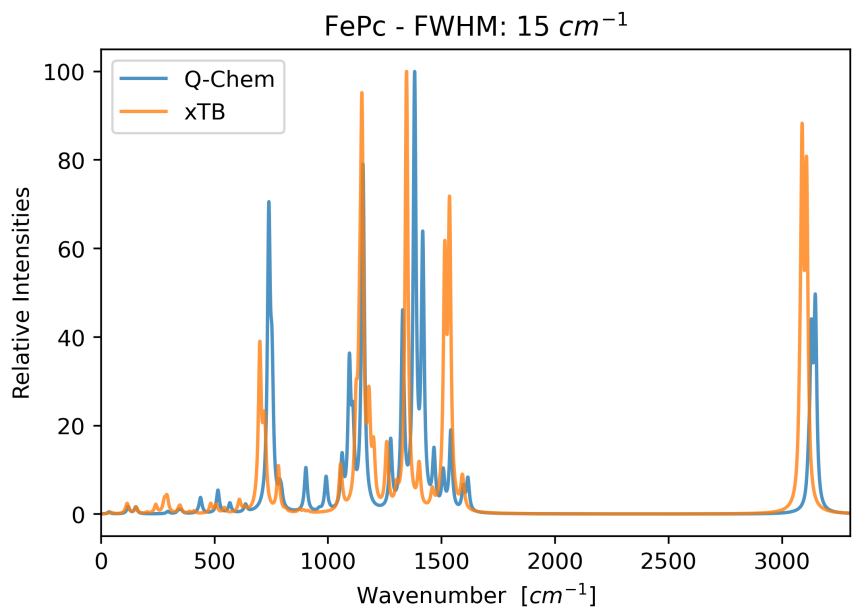

Figure S4: IR adsorption spectra of FePc obtained from GFN-xTB and Q-Chem calculations (see main article). Vibrations above 3000 cm$^{-1}$ correspond to motions of H atoms; the larger peaks around 1000 cm$^{-1}$ are mostly planar motions of N atoms. A Lorenz line shape function with FWHM of 10 cm$^{-1}$ was applied.

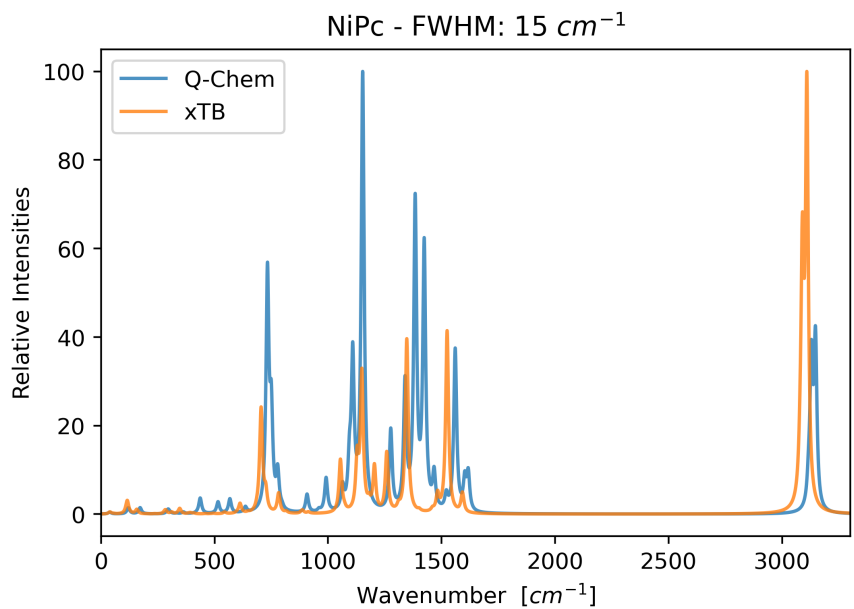

Figure S5: IR adsorption spectra of NiPc obtained from GFN-xTB and Q-Chem calculations (see main article). Vibrations above 3000 cm$^{-1}$ correspond to motions of H atoms; the larger peaks around 1000 cm$^{-1}$ are mostly planar motions of N atoms. A Lorenz line shape function with FWHM of 10 cm$^{-1}$ was applied.

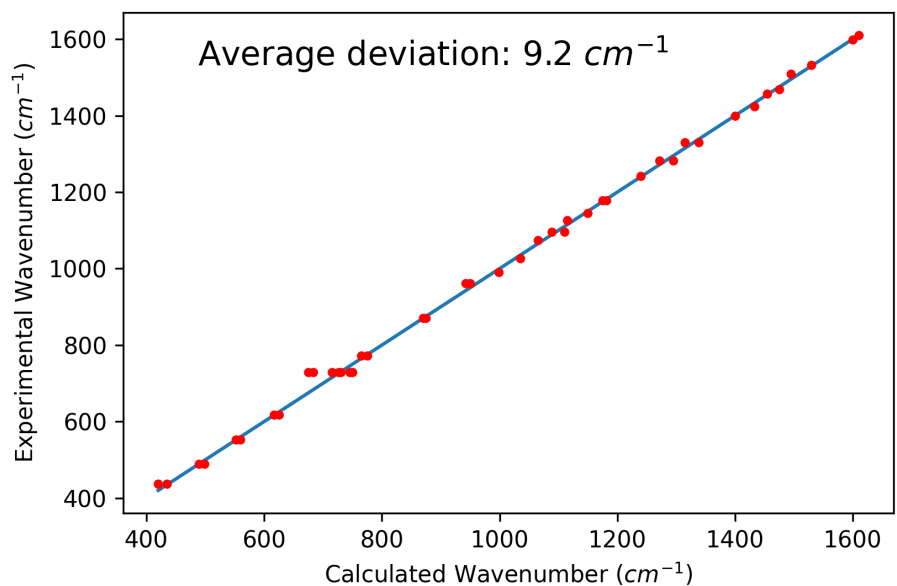

Figure S6: DFT-calculated vs experimental wavenumber for H$_2$Pc using a DFT calculation. A scale factor of 0.998 was applied to the calculated spectrum.

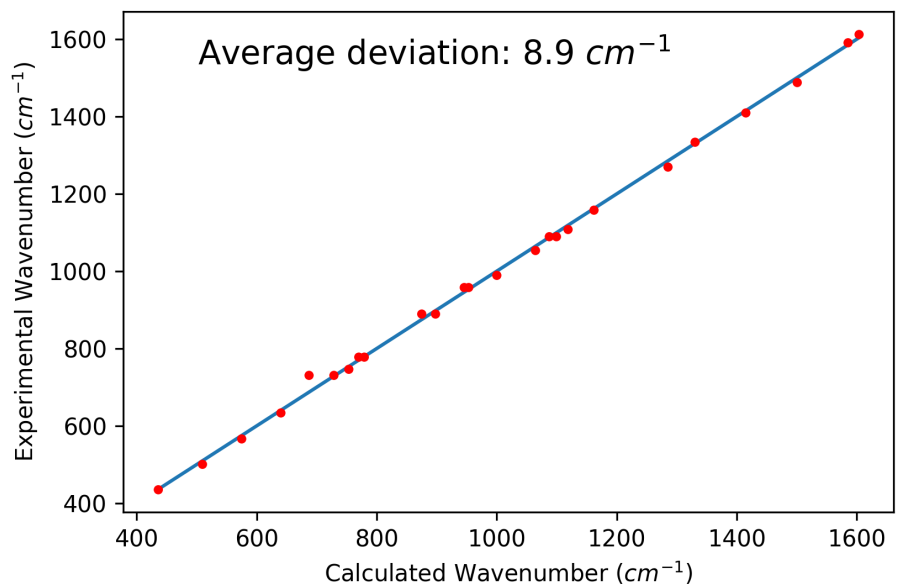

Figure S7: DFT-calculated vs experimental wavenumber for CuPc using a DFT calculation. A scale factor of 0.998 was applied to the calculated spectrum.

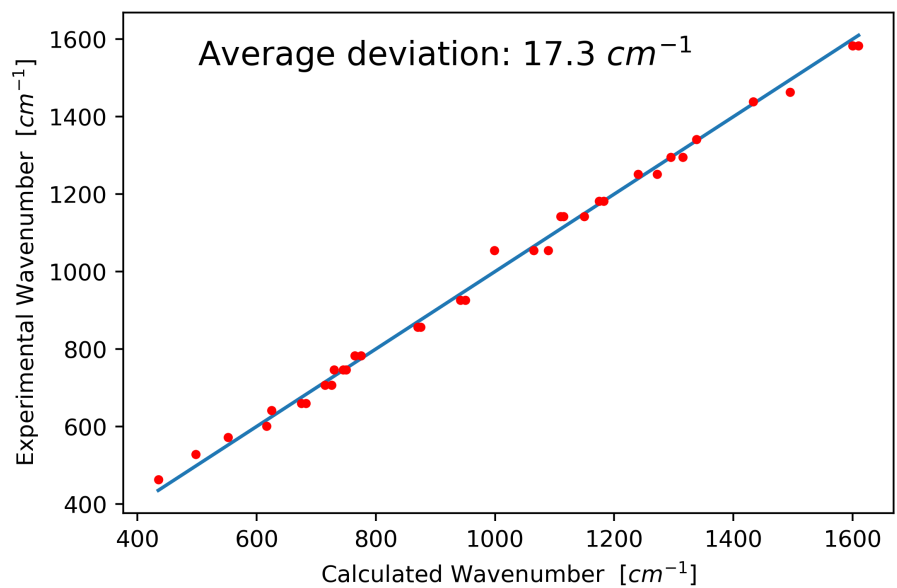

Figure S8: xTB-calculated vs experimental wavenumber for $H_2Pc$ using a xTB calculation.

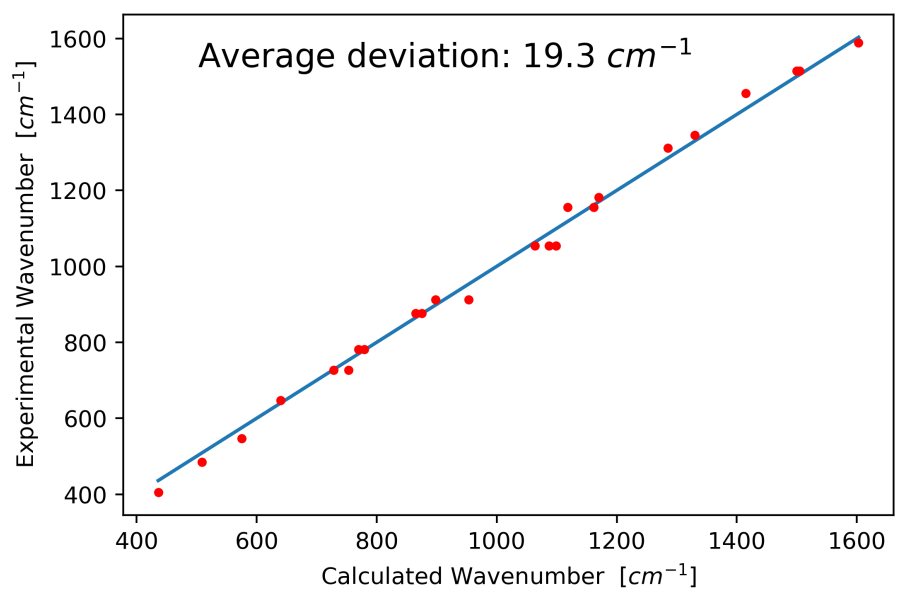

Figure S9: Calculated vs experimental wavenumber for CuPc using a xTB calculation.

# S4 Electron density plots

This section contains electron density plots as they occur during pseudorotation, calculated via a DFT-approach in Q-Chem with a b97d GGA functional and def2-SVP basis set. All plots assumed a vibrational amplitude according to the standard deviation of the first excited state of the corresponding harmonic oscillator. The eigenvectors are not plotted to scale, but are enlarged.

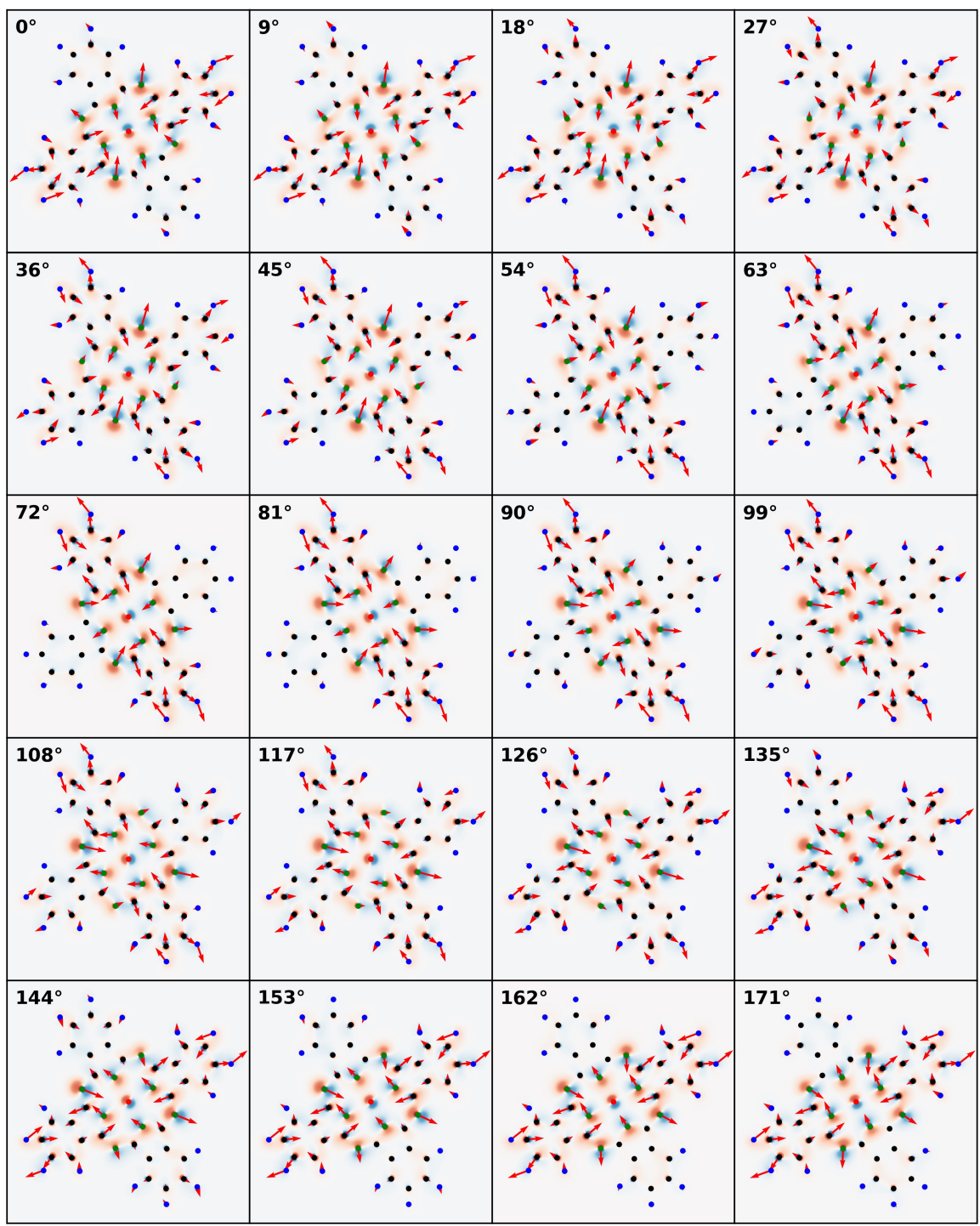

Figure S10: Change of electron density during vibration for the 891 cm$^{-1}$ mode of CuPc. The plots are made in 9 degree steps in terms of the phase angle $\omega t$. The electron density has been integrated along the z-direction and is plotted on a logarithmic scale. Excess and lack of electron density is denoted as blue or red. Distances in Bohr. Red arrows indicate the eigenvectors.

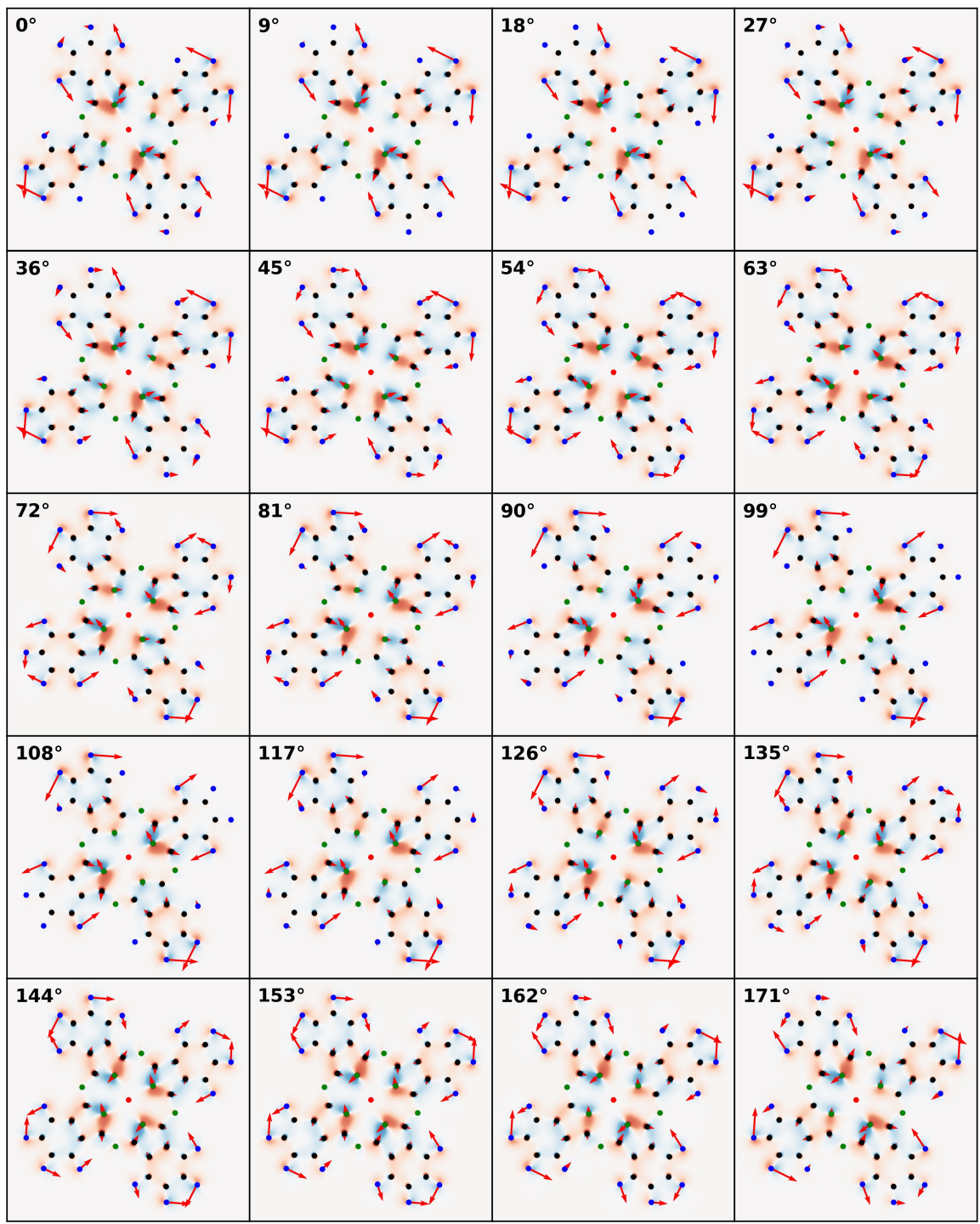

Figure S11: Change of electron density during vibration for the 1160 cm$^{-1}$ mode of CuPc. The plots are made in 9 degree steps in terms of the phase angle $\omega t$. The electron density has been integrated along the z-direction and is plotted on a logarithmic scale. Excess and lack of electron density is denoted as blue or red. Distances in Bohr. Red arrows indicate the eigenvectors.

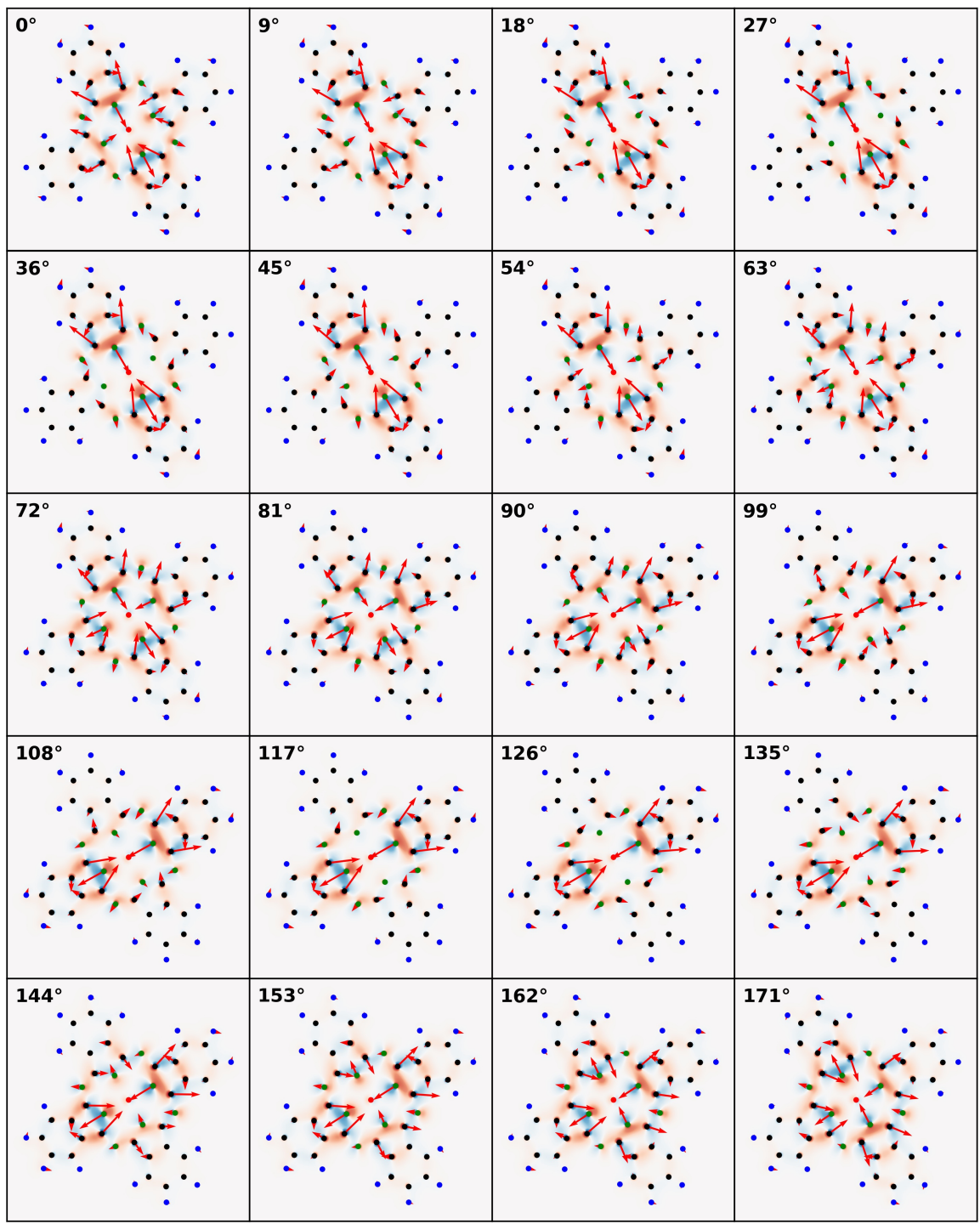

Figure S12: Change of electron density during vibration for the 1336 cm$^{-1}$ mode of CuPc. The plots are made in 9 degree steps in terms of the phase angle $\omega t$. The electron density has been integrated along the z-direction and is plotted on a logarithmic scale. Excess and lack of electron density is denoted as blue or red. Distances in Bohr. Red arrows indicate the eigenvectors.

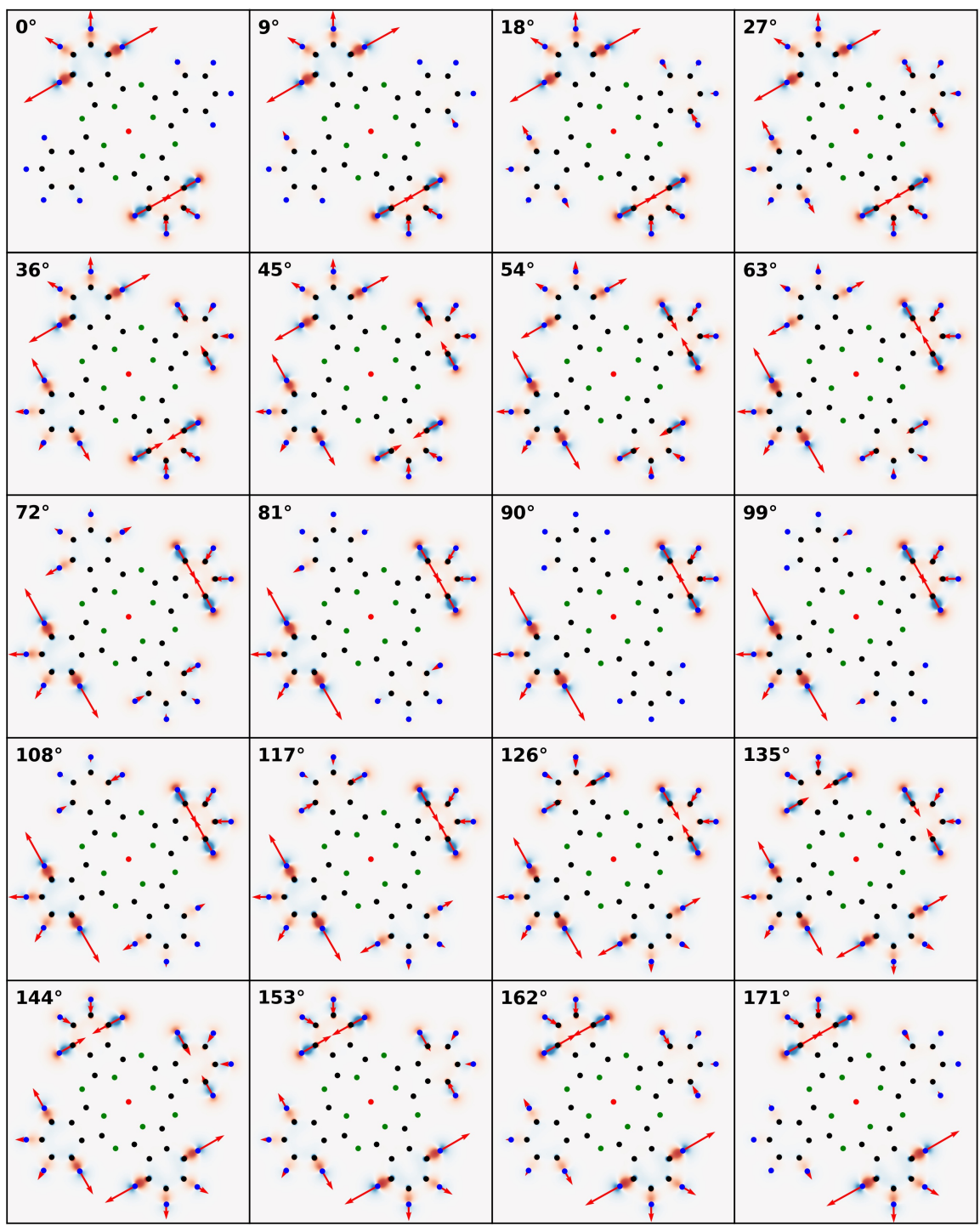

Figure S13: Change of electron density during vibration for the 3144 cm$^{-1}$ mode of CuPc. The plots are made in 9 degree steps in terms of the phase angle $\omega t$. The electron density has been integrated along the z-direction and is plotted on a logarithmic scale. Excess and lack of electron density is denoted as blue or red. Distances in Bohr. Red arrows indicate the eigenvectors.

# S5 Electric dipole moment

## S5.1 H$_2$Pc dipole scans

Pseudorotational excitation translates into a rotating electric dipole, as can be seen from Figure S14 below, featuring a phase difference of 90° between its two Cartesian components in the molecular plane. Note the slight deviation in amplitude in case of H$_2$Pc due to the reduced symmetry of the molecular system.

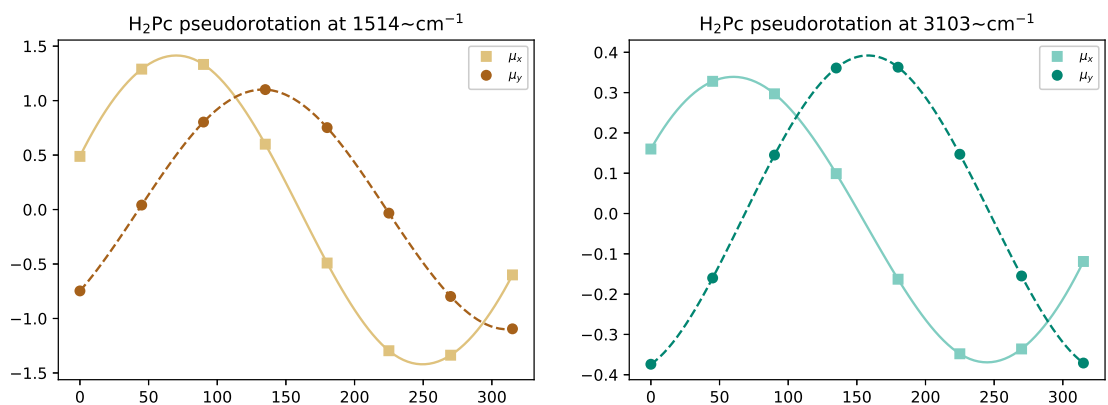

Figure S14: Dipole scans for H$_2$Pc...Scan over the electric dipole moment of H$_2$Pc as a function of the phase angle $\phi = \omega t$, corresponding to vibrational excitations at 1526 and 3144 cm$^{-1}$, calculated with DFT ansatz.

## S5.2 Rotating dipoles interpreted via the center of charge

The following Figures show how to imagine the rotating electric dipole as arising from the change in the center of charge. As the electron density and nuclei change during vibration, also the positive and negative centers of charge change. The distance from the positive to the negative center of charge gives the direction of the electric dipole moment.

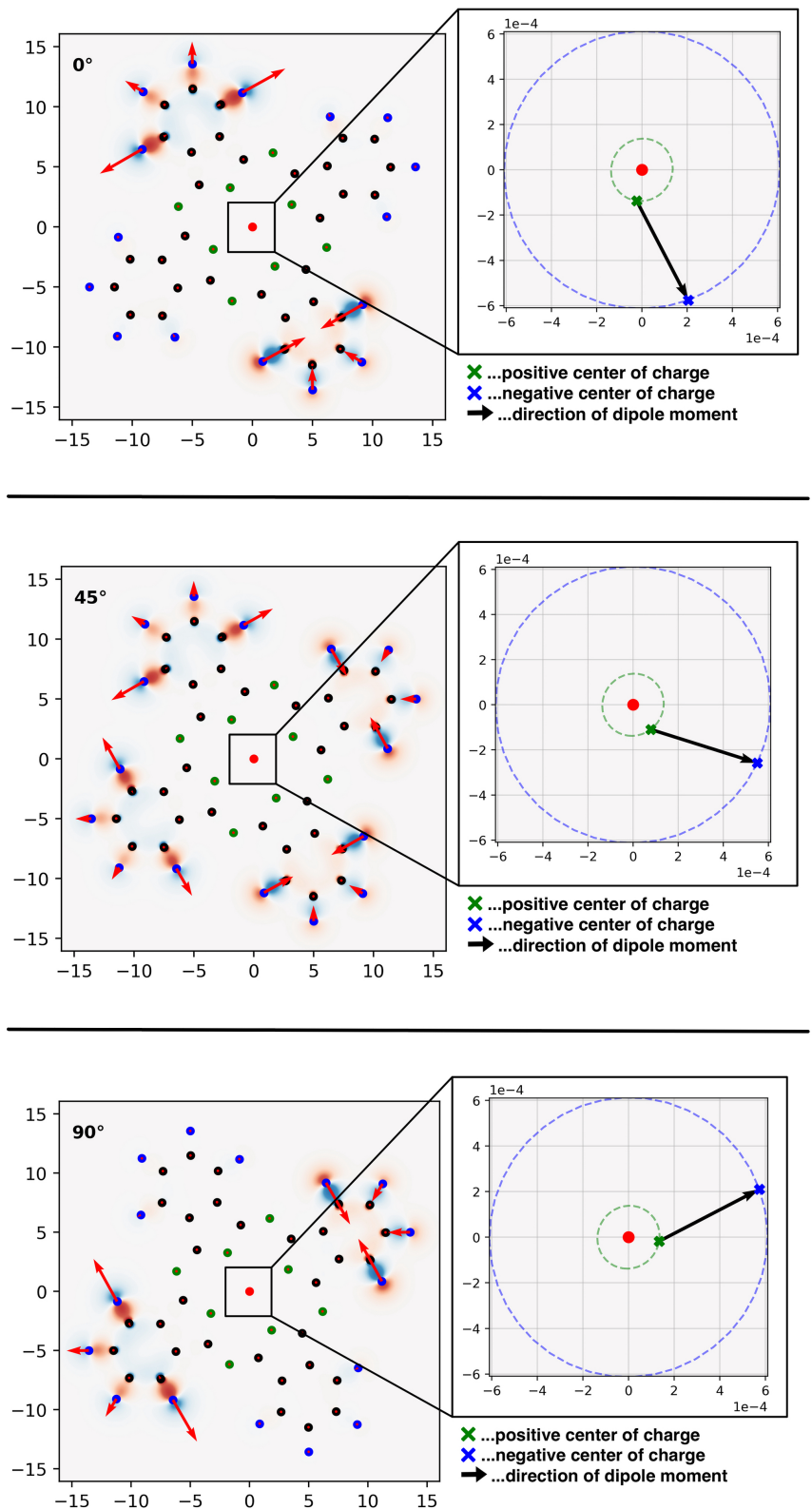

Figure S15: Change of electron density during vibration for the 3144 cm$^{-1}$ mode of CuPc for different phase angles $\omega t$ (left). Positive and negative centers of charge and direction of the dipole moment (right).

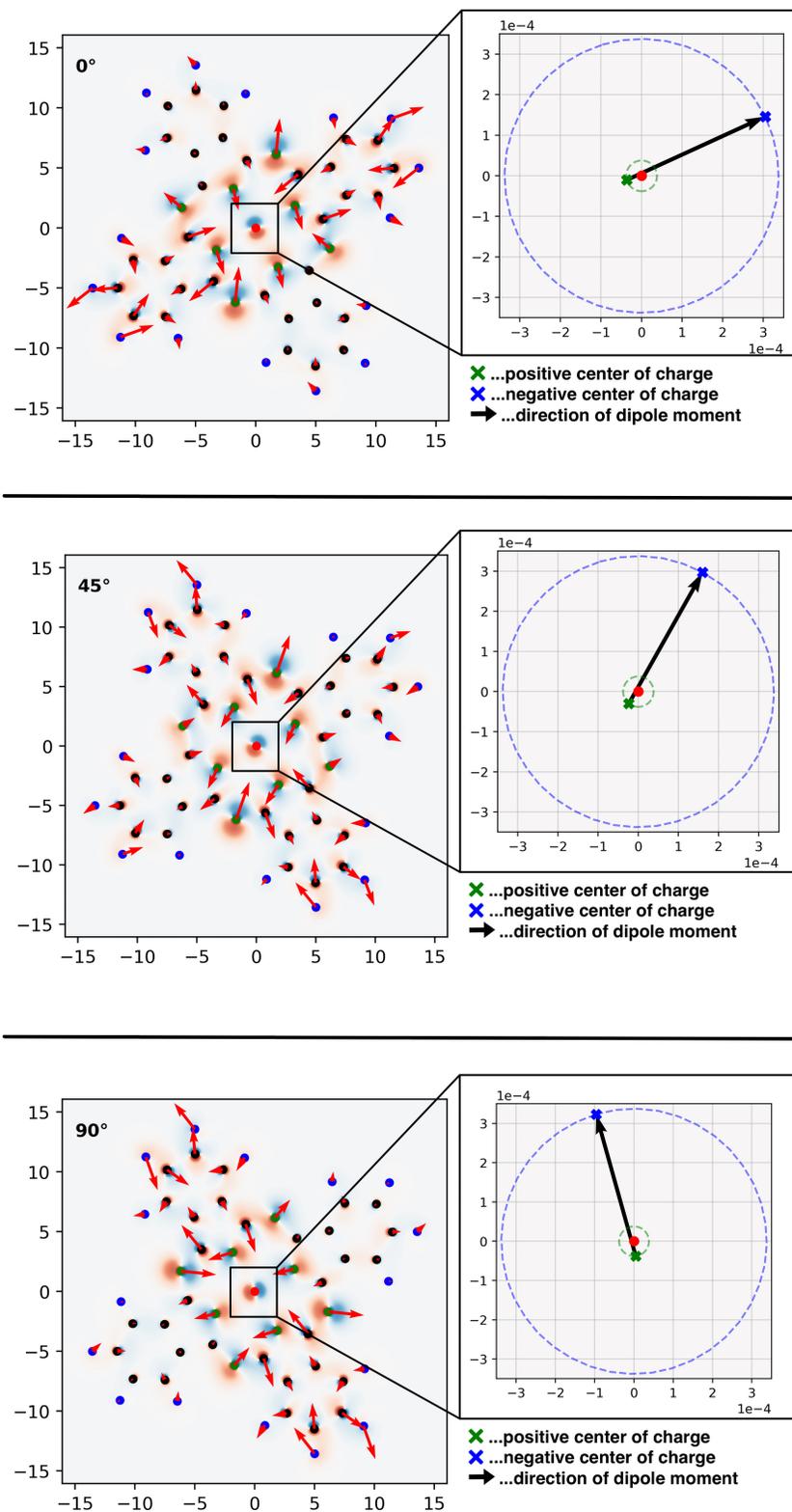

Figure S16: Change of electron density during vibration for the 891 cm$^{-1}$ mode of CuPc for different phase angles $\omega t$ (left). Positive and negative centers of charge and direction of the dipole moment (right).

## S5.3 Rotating dipoles calculated from partial charges

Figure S17 illustrates how partial charges can reproduce the electric dipole moment. In the left figure, the partial charges were calculated in the equilibrium position of the atoms (i.e. no displacement due to vibration was assumed). Then, the atomic sites where the partial charges sit were displaced from their equilibrium position according to the eigenvectors of vibration. As expected, these equilibrium partial charges do not reproduce the electric dipole moment at all. Only the APT charge, which is independent of the displacement, does reproduce the electric dipole moment.

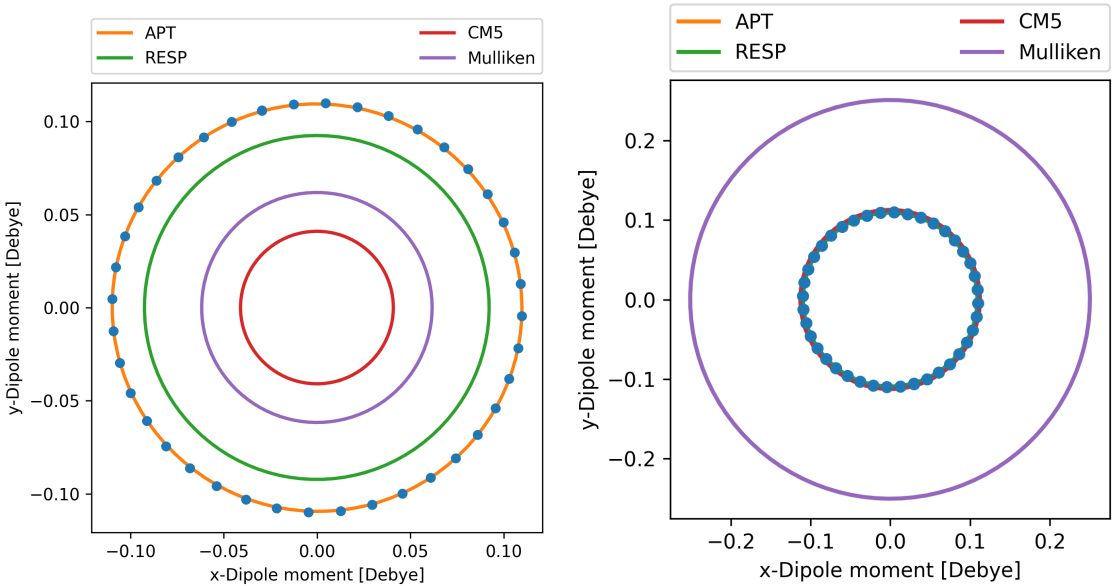

Figure S17: Electric dipole moment calculated via the use of different partial charge models for the 891 cm$^{-1}$ mode. Blue points denote the values calculated directly by Q-Chem via DFT at different phase angles $\phi = \omega t$. Left: Partial charges were calculated for the undisturbed equilibrium geometry of the molecule. Right: Partial charges were calculated assuming a distorted molecule according to the vibrational eigenvectors. In the right plot all charges, except Mulliken partial charges, reproduce the calculated dipole moment from Q-Chem and therefore overlap.

Re-evaluating, however, the partial charges at the displaced atomic sites at every phase angle $\omega t$ leads to much better results. In the right figure, the dipole moment produced by these re-evaluated partial charges are plotted. Indeed, all but Mulliken charges reproduce the electric dipole moment within a few percentage points. Nevertheless, it has to be noted that CM5 charges use an semi-empirical method for calculating partial charges. This semi-empirical method does not necessarily conserve charge neutrality and therefore the dipole moment. On the other hand, RESP charges are often calculated under the constraint of reproducing the electric dipole moment. Q-Chem, however, does not do this and small deviations from the exact dipole moment are expected and observed.

# S6 Optimized Geometries

This section provides the optimized geometries of all the neutral phathalocyanines treated in the main article (H$_2$Pc, FePc, CoPc, NiPc, CuPc). These structures were calculated with an DFT approch using a b97d GGA functional and def2-SVP basis in Q-Chem The structures are given in Cartesian coordinates in units of Ångstrom. The formatting supports direct copy and paste from the pdf document.

## S6.1 H$_2$Pc

```
58

C -0.70681 -4.19123 0.00000

C -1.43500 -5.38904 0.00000

C -0.70736 -6.59405 0.00000

C 0.70736 -6.59405 0.00000

C 1.43500 -5.38904 0.00000

C 0.70681 -4.19123 0.00000

H -1.24278 -7.55224 0.00000

H 1.24278 -7.55224 0.00000

H 2.53157 -5.37665 0.00000

H -2.53157 -5.37665 0.00000

C 1.10249 -2.77246 0.00000

N -0.00000 -1.96175 0.00000

C -1.10249 -2.77246 0.00000

N -2.39045 -2.40438 0.00000

C -6.60236 -0.70950 0.00000

C -6.60236 0.70950 0.00000

N -2.04175 0.00000 0.00000
```

```
C -5.40309 -1.43640 0.00000

C -5.40309 1.43640 0.00000

C -4.19858 -0.71157 0.00000

C -4.19858 0.71157 0.00000

H -7.56097 -1.24364 0.00000

H -7.56097 1.24364 0.00000

C -2.80753 -1.15161 0.00000

C -2.80753 1.15161 0.00000

H -5.38931 -2.53256 0.00000

H -5.38931 2.53256 0.00000

N -2.39045 2.40438 0.00000

C -0.70736 6.59405 0.00000

N 2.04175 0.00000 0.00000

C 0.70736 6.59405 0.00000

N -0.00000 1.96175 0.00000

C -1.43500 5.38904 0.00000

C 1.43500 5.38904 0.00000

C -0.70681 4.19123 0.00000

C 0.70681 4.19123 0.00000

H -1.24278 7.55224 0.00000

H 1.24278 7.55224 0.00000

C -1.10249 2.77246 0.00000

C 1.10249 2.77246 0.00000

H -2.53157 5.37665 0.00000

C 2.80753 1.15161 0.00000

H 2.53157 5.37665 0.00000

C 2.80753 -1.15161 0.00000

C 4.19858 0.71157 0.00000

C 4.19858 -0.71157 0.00000
```

```
C 5.40309 1.43640 0.00000

N 2.39045 2.40438 0.00000

C 5.40309 -1.43640 0.00000

N 2.39045 -2.40438 0.00000

C 6.60236 0.70950 0.00000

C 6.60236 -0.70950 0.00000

H 5.38931 2.53256 0.00000

H 5.38931 -2.53256 0.00000

H 7.56097 1.24364 0.00000

H 7.56097 -1.24364 0.00000

H 1.02048 0.00000 0.00000

H -1.02048 0.00000 0.00000
```

## S6.2  FePc

```
57

C 0.70746 4.16643 -0.00000

C 1.43676 5.36721 0.00000

C 0.70922 6.56789 0.00000

C -0.70922 6.56789 0.00000

C -1.43676 5.36721 -0.00000

C -0.70746 4.16643 -0.00000

H 1.24332 7.52667 0.00000

H -1.24332 7.52667 0.00000

H -2.53299 5.35289 -0.00000

H 2.53299 5.35289 0.00000

C -1.12030 2.76668 -0.00000

N -0.00000 1.95216 -0.00000

C 1.12030 2.76668 -0.00000
```

```
N 2.39139 2.39139 -0.00000
C 6.56789 0.70922 -0.00000
C 6.56789 -0.70922 0.00000
N 1.95216 0.00000 0.00000
C 5.36721 1.43676 -0.00000
C 5.36721 -1.43676 0.00000
C 4.16643 0.70746 -0.00000
C 4.16643 -0.70746 0.00000
H 7.52667 1.24332 -0.00000
H 7.52667 -1.24332 0.00000
C 2.76668 1.12030 -0.00000
C 2.76668 -1.12030 0.00000
H 5.35289 2.53299 -0.00000
H 5.35289 -2.53299 0.00000
N 2.39139 -2.39139 0.00000
C 0.70922 -6.56789 -0.00000
N -1.95216 0.00000 -0.00000
C -0.70922 -6.56789 -0.00000
N -0.00000 -1.95216 0.00000
C 1.43676 -5.36721 0.00000
C -1.43676 -5.36721 -0.00000
C 0.70746 -4.16643 0.00000
C -0.70746 -4.16643 0.00000
H 1.24332 -7.52667 -0.00000
H -1.24332 -7.52667 0.00000
C 1.12030 -2.76668 0.00000
C -1.12030 -2.76668 0.00000
H 2.53299 -5.35289 0.00000
C -2.76668 -1.12030 0.00000
```

```
H -2.53299 -5.35289 -0.00000
C -2.76668 1.12030 -0.00000
C -4.16642 -0.70746 0.00000
C -4.16642 0.70746 -0.00000
C -5.36720 -1.43676 0.00000
N -2.39139 -2.39139 0.00000
C -5.36720 1.43676 -0.00000
N -2.39139 2.39139 -0.00000
C -6.56789 -0.70922 0.00000
C -6.56789 0.70922 -0.00000
H -5.35289 -2.53299 0.00000
H -5.35289 2.53299 -0.00000
H -7.52667 -1.24332 0.00000
H -7.52667 1.24332 -0.00000
Fe -0.00000 0.00000 0.00000
```

## S6.3 CoPc

```
57

C 0.70747 4.15212 0.00000
C 1.43803 5.35429 0.00000
C 0.71048 6.55255 0.00000
C -0.71048 6.55255 0.00000
C -1.43803 5.35429 0.00000
C -0.70747 4.15212 0.00000
H 1.24364 7.51185 0.00000
H -1.24364 7.51185 0.00000
H -2.53414 5.33981 0.00000
H 2.53414 5.33981 0.00000
```

```
C -1.11957 2.75776 0.00000
N 0.00000 1.93766 0.00000
C 1.11957 2.75776 0.00000
N 2.38981 2.38693 0.00000
C 6.55555 0.70832 0.00000
C 6.55555 -0.70832 0.00000
N 1.92852 0.00000 0.00000
C 5.35293 1.43595 0.00000
C 5.35293 -1.43595 0.00000
C 4.15466 0.70516 0.00000
C 4.15466 -0.70516 0.00000
H 7.51378 1.24339 0.00000
H 7.51378 -1.24339 0.00000
C 2.75333 1.11713 0.00000
C 2.75333 -1.11713 0.00000
H 5.33796 2.53215 0.00000
H 5.33796 -2.53215 0.00000
N 2.38981 -2.38693 0.00000
C 0.71048 -6.55255 0.00000
N -1.92852 0.00000 0.00000
C -0.71048 -6.55255 0.00000
N 0.00000 -1.93766 0.00000
C 1.43803 -5.35429 0.00000
C -1.43803 -5.35429 0.00000
C 0.70747 -4.15212 0.00000
C -0.70747 -4.15212 0.00000
H 1.24364 -7.51185 0.00000
H -1.24364 -7.51185 0.00000
C 1.11957 -2.75776 0.00000
```

```
C -1.11957 -2.75776 0.00000
H 2.53414 -5.33981 0.00000
C -2.75333 -1.11713 0.00000
H -2.53414 -5.33981 0.00000
C -2.75333 1.11713 0.00000
C -4.15466 -0.70516 0.00000
C -4.15466 0.70516 0.00000
C -5.35293 -1.43595 0.00000
N -2.38981 -2.38693 0.00000
C -5.35293 1.43595 0.00000
N -2.38981 2.38693 0.00000
C -6.55555 -0.70832 0.00000
C -6.55555 0.70832 0.00000
H -5.33796 -2.53215 0.00000
H -5.33796 2.53215 0.00000
H -7.51378 -1.24339 0.00000
H -7.51378 1.24339 0.00000
Co 0.00000 0.00000 0.00000
```

## S6.4  NiPc

```
57

C 0.70517 4.14776 0.00000
C 1.43690 5.34678 0.00000
C 0.70910 6.54754 0.00000
C -0.70910 6.54754 0.00000
C -1.43690 5.34678 0.00000
C -0.70517 4.14776 0.00000
H 1.24340 7.50619 0.00000
```

```
H -1.24340 7.50619 0.00000
H -2.53305 5.33175 0.00000
H 2.53305 5.33175 0.00000
C -1.11602 2.74864 0.00000
N -0.00000 1.92621 0.00000
C 1.11602 2.74864 0.00000
N 2.38574 2.38574 0.00000
C 6.54754 0.70910 0.00000
C 6.54754 -0.70910 0.00000
N 1.92621 0.00000 0.00000
C 5.34678 1.43690 0.00000
C 5.34678 -1.43690 0.00000
C 4.14776 0.70517 0.00000
C 4.14776 -0.70517 0.00000
H 7.50619 1.24340 0.00000
H 7.50619 -1.24340 0.00000
C 2.74864 1.11602 0.00000
C 2.74865 -1.11602 0.00000
H 5.33175 2.53305 0.00000
H 5.33175 -2.53305 0.00000
N 2.38574 -2.38574 0.00000
C 0.70910 -6.54754 0.00000
N -1.92621 0.00000 0.00000
C -0.70910 -6.54754 0.00000
N -0.00000 -1.92621 0.00000
C 1.43690 -5.34678 0.00000
C -1.43690 -5.34678 0.00000
C 0.70517 -4.14776 0.00000
C -0.70517 -4.14776 0.00000
```

```
H 1.24340 -7.50619 0.00000
H -1.24340 -7.50619 0.00000
C 1.11602 -2.74864 0.00000
C -1.11602 -2.74864 0.00000
H 2.53305 -5.33175 0.00000
C -2.74864 -1.11602 0.00000
H -2.53305 -5.33175 0.00000
C -2.74864 1.11602 0.00000
C -4.14776 -0.70517 0.00000
C -4.14776 0.70517 0.00000
C -5.34678 -1.43690 0.00000
N -2.38574 -2.38574 0.00000
C -5.34678 1.43690 0.00000
N -2.38574 2.38574 0.00000
C -6.54754 -0.70910 0.00000
C -6.54754 0.70910 0.00000
H -5.33175 -2.53305 0.00000
H -5.33175 2.53305 0.00000
H -7.50619 -1.24340 0.00000
H -7.50619 1.24340 0.00000
Ni -0.00000 0.00000 0.00000
```

## S6.5  CuPc

```
57

C 0.70853 4.18888 0.00000
C 1.43593 5.38899 0.00000
C 0.70867 6.59106 0.00000
C -0.70867 6.59106 -0.00000
```

```
C -1.43593 5.38899 -0.00000
C -0.70853 4.18888 -0.00000
H 1.24389 7.54924 0.00000
H -1.24389 7.54924 -0.00000
H -2.53205 5.37306 -0.00000
H 2.53205 5.37306 0.00000
C -1.12471 2.78402 -0.00000
N 0.00000 1.98843 0.00000
C 1.12471 2.78402 0.00000
N 2.39540 2.39540 0.00000
C 6.59106 0.70867 0.00000
C 6.59106 -0.70867 0.00000
N 1.98843 -0.00000 0.00000
C 5.38899 1.43593 0.00000
C 5.38899 -1.43593 0.00000
C 4.18888 0.70853 0.00000
C 4.18888 -0.70853 0.00000
H 7.54924 1.24389 0.00000
H 7.54924 -1.24389 0.00000
C 2.78402 1.12471 0.00000
C 2.78402 -1.12471 0.00000
H 5.37306 2.53205 0.00000
H 5.37306 -2.53205 0.00000
N 2.39540 -2.39540 0.00000
C 0.70867 -6.59106 0.00000
N -1.98843 0.00000 -0.00000
C -0.70867 -6.59106 -0.00000
N -0.00000 -1.98843 -0.00000
C 1.43593 -5.38899 0.00000
```

```
C -1.43593 -5.38899 -0.00000
C 0.70853 -4.18888 0.00000
C -0.70853 -4.18888 -0.00000
H 1.24389 -7.54924 0.00000
H -1.24389 -7.54924 -0.00000
C 1.12471 -2.78402 0.00000
C -1.12471 -2.78402 -0.00000
H 2.53205 -5.37306 0.00000
C -2.78402 -1.12471 -0.00000
H -2.53205 -5.37306 -0.00000
C -2.78402 1.12471 -0.00000
C -4.18888 -0.70853 -0.00000
C -4.18888 0.70853 -0.00000
C -5.38899 -1.43593 -0.00000
N -2.39540 -2.39540 -0.00000
C -5.38899 1.43593 -0.00000
N -2.39540 2.39540 -0.00000
C -6.59106 -0.70867 -0.00000
C -6.59106 0.70867 -0.00000
H -5.37306 -2.53205 -0.00000
H -5.37306 2.53205 -0.00000
H -7.54924 -1.24389 -0.00000
H -7.54924 1.24389 -0.00000
Cu 0.00000 0.00000 0.00000
```

# S7 Vibrational Eigenmodes

This section contains excerpts of the CoPc output files obtained with the Q-Chem program package as used for the calculation of vibrational g-factors. It provides the equilibrium positions, some vibrational frequencies and their corresponding eigenmodes.

## S7.1 891 cm$^{-1}$

Table S3: Eigenvectors for the 891 cm$^{-1}$ eigenmode of CuPc. Each eigenvector is normalized to one and not mass-weighted. $\vec{a}_i$ and $\vec{b}_i$ correspond to the eigenvector components pointing into one of the two planar Cartesian coordinate directions. The positions of the atoms in space are denoted as $x$, $y$, $z$ with units in Angstrom.

| Atom | x | y | z | $\vec{a}_x$ | $\vec{a}_y$ | $\vec{a}_z$ | $\vec{b}_x$ | $\vec{b}_y$ | $\vec{b}_z$ |
|---|---|---|---|---|---|---|---|---|---|
| C | 0.708 | 4.188 | 0.000 | 0.001 | 0.085 | 0.000 | -0.003 | -0.002 | -0.000 |
| C | 1.436 | 5.389 | 0.000 | 0.088 | -0.041 | -0.000 | 0.040 | 0.015 | 0.000 |
| C | 0.708 | 6.591 | 0.000 | -0.081 | -0.173 | -0.000 | 0.009 | 0.007 | 0.000 |
| C | -0.708 | 6.591 | 0.000 | -0.080 | 0.172 | 0.000 | -0.015 | 0.022 | -0.000 |
| C | -1.436 | 5.389 | 0.000 | 0.092 | 0.039 | 0.000 | -0.032 | 0.018 | -0.000 |
| C | -0.708 | 4.188 | 0.000 | 0.000 | -0.085 | -0.000 | 0.003 | -0.009 | 0.000 |
| H | 1.243 | 7.549 | 0.000 | 0.041 | -0.243 | 0.000 | -0.009 | 0.018 | -0.000 |
| H | -1.243 | 7.549 | 0.000 | 0.040 | 0.241 | -0.000 | 0.012 | 0.039 | 0.000 |
| H | -2.532 | 5.376 | 0.000 | 0.090 | -0.047 | -0.000 | -0.034 | 0.027 | 0.000 |
| H | 2.532 | 5.376 | 0.000 | 0.086 | 0.044 | 0.000 | 0.041 | 0.031 | -0.000 |
| C | -1.124 | 2.784 | 0.000 | 0.039 | -0.216 | 0.000 | -0.001 | -0.027 | -0.000 |
| N | 0.000 | 1.988 | 0.000 | 0.181 | 0.007 | -0.000 | 0.008 | -0.153 | -0.000 |
| C | 1.124 | 2.784 | 0.000 | 0.039 | 0.217 | 0.000 | 0.005 | -0.008 | 0.000 |
| N | 2.395 | 2.395 | 0.000 | -0.188 | -0.111 | -0.000 | 0.096 | 0.179 | -0.000 |
| C | 6.591 | 0.708 | 0.000 | -0.022 | -0.015 | -0.000 | 0.172 | 0.080 | -0.000 |
| C | 6.591 | -0.708 | 0.000 | -0.007 | 0.009 | 0.000 | -0.173 | 0.081 | 0.000 |
| N | 1.988 | 0.000 | 0.000 | 0.153 | 0.008 | 0.000 | 0.007 | -0.181 | -0.000 |
| C | 5.389 | 1.436 | 0.000 | -0.018 | -0.032 | -0.000 | 0.039 | -0.092 | -0.000 |
| C | 5.389 | -1.436 | 0.000 | -0.015 | 0.040 | 0.000 | -0.041 | -0.088 | 0.000 |
| C | 4.188 | 0.708 | 0.000 | 0.009 | 0.003 | 0.000 | -0.085 | -0.000 | 0.000 |
| C | 4.188 | -0.708 | 0.000 | 0.002 | -0.003 | -0.000 | 0.085 | -0.001 | -0.000 |
| H | 7.549 | 1.243 | 0.000 | -0.039 | 0.012 | 0.000 | 0.241 | -0.040 | 0.000 |
| H | 7.549 | -1.243 | 0.000 | -0.018 | -0.009 | -0.000 | -0.243 | -0.041 | -0.000 |
| C | 2.784 | 1.124 | 0.000 | 0.027 | -0.001 | -0.000 | -0.216 | -0.039 | -0.000 |
| C | 2.784 | -1.124 | 0.000 | 0.008 | 0.005 | 0.000 | 0.217 | -0.039 | 0.000 |
| H | 5.376 | 2.532 | 0.000 | -0.027 | -0.034 | 0.000 | -0.047 | -0.090 | 0.000 |
| H | 5.376 | -2.532 | 0.000 | -0.031 | 0.041 | -0.000 | 0.044 | -0.086 | -0.000 |
| N | 2.395 | -2.395 | 0.000 | -0.179 | 0.096 | -0.000 | -0.111 | 0.188 | -0.000 |
| C | 0.708 | -6.591 | 0.000 | -0.080 | 0.172 | 0.000 | -0.015 | 0.022 | 0.000 |
| N | -1.988 | -0.000 | 0.000 | 0.153 | 0.008 | 0.000 | 0.007 | -0.181 | 0.000 |
| C | -0.708 | -6.591 | 0.000 | -0.081 | -0.173 | -0.000 | 0.009 | 0.007 | -0.000 |
| N | -0.000 | -1.988 | 0.000 | 0.181 | 0.007 | 0.000 | 0.008 | -0.153 | 0.000 |
| C | 1.436 | -5.389 | 0.000 | 0.092 | 0.039 | 0.000 | -0.032 | 0.018 | 0.000 |
| C | -1.436 | -5.389 | 0.000 | 0.088 | -0.041 | -0.000 | 0.040 | 0.015 | -0.000 |
| C | 0.708 | -4.188 | 0.000 | 0.000 | -0.085 | -0.000 | 0.003 | -0.009 | -0.000 |
| C | -0.708 | -4.188 | 0.000 | 0.001 | 0.085 | 0.000 | -0.003 | -0.002 | 0.000 |
| H | 1.243 | -7.549 | 0.000 | 0.040 | 0.241 | -0.000 | 0.012 | 0.039 | -0.000 |
| H | -1.243 | -7.549 | 0.000 | 0.041 | -0.243 | 0.000 | -0.009 | 0.018 | 0.000 |
| C | 1.124 | -2.784 | 0.000 | 0.039 | -0.216 | 0.000 | -0.001 | -0.027 | 0.000 |
| C | -1.124 | -2.784 | 0.000 | 0.039 | 0.217 | -0.000 | 0.005 | -0.008 | -0.000 |
| H | 2.532 | -5.376 | 0.000 | 0.090 | -0.047 | -0.000 | -0.034 | 0.027 | -0.000 |
| C | -2.784 | -1.124 | 0.000 | 0.027 | -0.001 | -0.000 | -0.216 | -0.039 | -0.000 |
| H | -2.532 | -5.376 | 0.000 | 0.086 | 0.044 | 0.000 | 0.041 | 0.031 | 0.000 |
| C | -2.784 | 1.124 | 0.000 | 0.008 | 0.005 | 0.000 | 0.217 | -0.039 | 0.000 |
| C | -4.188 | -0.708 | 0.000 | 0.009 | 0.003 | 0.000 | -0.085 | -0.000 | 0.000 |
| C | -4.188 | 0.708 | 0.000 | 0.002 | -0.003 | -0.000 | 0.085 | -0.001 | -0.000 |
| C | -5.389 | -1.436 | 0.000 | -0.018 | -0.032 | -0.000 | 0.039 | -0.092 | -0.000 |
| N | -2.395 | -2.395 | 0.000 | -0.188 | -0.111 | 0.000 | 0.096 | 0.179 | 0.000 |
| C | -5.389 | 1.436 | 0.000 | -0.015 | 0.040 | 0.000 | -0.041 | -0.088 | 0.000 |
| N | -2.395 | 2.395 | 0.000 | -0.179 | 0.096 | -0.000 | -0.111 | 0.188 | -0.000 |
| C | -6.591 | -0.708 | 0.000 | -0.022 | -0.015 | -0.000 | 0.172 | 0.080 | -0.000 |
| C | -6.591 | 0.708 | 0.000 | -0.007 | 0.009 | 0.000 | -0.173 | 0.081 | 0.000 |
| H | -5.376 | -2.532 | 0.000 | -0.027 | -0.034 | -0.000 | -0.047 | -0.090 | 0.000 |
| H | -5.376 | 2.532 | 0.000 | -0.031 | 0.041 | -0.000 | 0.044 | -0.086 | -0.000 |
| H | -7.549 | -1.243 | 0.000 | -0.039 | 0.012 | 0.000 | 0.241 | -0.040 | 0.000 |
| H | -7.549 | 1.243 | 0.000 | -0.018 | -0.009 | -0.000 | -0.243 | -0.041 | -0.000 |
| Cu | 0.000 | 0.000 | 0.000 | -0.022 | -0.001 | -0.000 | -0.001 | 0.022 | -0.000 |

## S7.2   1160 cm$^{-1}$

Table S4: Eigenvectors for the 1160 cm$^{-1}$ eigenmode of CuPc. Each eigenvector is normalized to one and not mass-weighted. $\vec{a}_i$ and $\vec{b}_i$ correspond to the eigenvector components pointing into one of the two planar Cartesian coordinate directions. The positions of the atoms in space are denoted as $x$, $y$, $z$ with units in Angstrom

| Atom | x | y | z | $\vec{a}_x$ | $\vec{a}_y$ | $\vec{a}_z$ | $\vec{b}_x$ | $\vec{b}_y$ | $\vec{b}_z$ |
|---|---|---|---|---|---|---|---|---|---|
| C | 0.708 | 4.188 | 0.00 | -0.043 | 0.073 | -0.000 | -0.029 | -0.027 | -0.000 |
| C | 1.436 | 5.389 | 0.00 | -0.011 | 0.014 | -0.000 | 0.028 | 0.010 | 0.000 |
| C | 0.708 | 6.591 | 0.00 | 0.017 | -0.035 | 0.000 | 0.005 | 0.008 | -0.000 |
| C | -0.708 | 6.591 | 0.00 | -0.013 | -0.035 | -0.000 | 0.012 | 0.010 | -0.000 |
| C | -1.436 | 5.389 | 0.00 | 0.023 | 0.008 | 0.000 | 0.020 | -0.015 | 0.000 |

| | | | | | | | | |
|---|---|---|---|---|---|---|---|---|---|
| C | -0.708 | 4.188 | 0.00 | 0.025 | 0.078 | -0.000 | -0.046 | -0.010 | -0.000 |
| H | 1.243 | 7.549 | 0.00 | 0.293 | -0.193 | -0.000 | -0.056 | 0.043 | 0.000 |
| H | -1.243 | 7.549 | 0.00 | -0.285 | -0.190 | -0.000 | 0.090 | 0.054 | 0.000 |
| H | -2.532 | 5.376 | 0.00 | 0.028 | -0.036 | -0.000 | 0.022 | -0.280 | -0.000 |
| H | 2.532 | 5.376 | 0.00 | -0.014 | 0.101 | 0.000 | 0.033 | 0.264 | 0.000 |
| C | -1.124 | 2.784 | 0.00 | -0.053 | 0.012 | 0.000 | -0.079 | 0.070 | 0.000 |
| N | 0.000 | 1.988 | 0.00 | 0.041 | -0.047 | -0.000 | 0.162 | 0.012 | 0.000 |
| C | 1.124 | 2.784 | 0.00 | 0.009 | -0.022 | 0.000 | -0.095 | -0.068 | -0.000 |
| N | 2.395 | 2.395 | 0.00 | -0.000 | -0.010 | -0.000 | -0.011 | 0.005 | -0.000 |
| C | 6.591 | 0.708 | 0.00 | -0.010 | 0.012 | -0.000 | -0.035 | 0.013 | 0.000 |
| C | 6.591 | -0.708 | 0.00 | -0.008 | 0.005 | -0.000 | -0.035 | -0.017 | 0.000 |
| N | 1.988 | 0.000 | 0.00 | -0.012 | 0.162 | -0.000 | -0.047 | -0.041 | 0.000 |
| C | 5.389 | 1.436 | 0.00 | 0.015 | 0.020 | 0.000 | 0.008 | -0.023 | -0.000 |
| C | 5.389 | -1.436 | 0.00 | -0.010 | 0.028 | 0.000 | 0.014 | 0.011 | -0.000 |
| C | 4.188 | 0.708 | 0.00 | 0.010 | -0.046 | -0.000 | 0.078 | -0.025 | 0.000 |
| C | 4.188 | -0.708 | 0.00 | 0.027 | -0.029 | -0.000 | 0.073 | 0.043 | 0.000 |
| H | 7.549 | 1.243 | 0.00 | -0.054 | 0.090 | 0.000 | -0.190 | 0.285 | -0.000 |
| H | 7.549 | -1.243 | 0.00 | -0.043 | -0.056 | 0.000 | -0.193 | -0.293 | -0.000 |
| C | 2.784 | 1.124 | 0.00 | -0.070 | -0.079 | 0.000 | 0.012 | 0.053 | -0.000 |
| C | 2.784 | -1.124 | 0.00 | 0.068 | -0.095 | 0.000 | -0.022 | -0.009 | 0.000 |
| H | 5.376 | 2.532 | 0.00 | 0.280 | 0.022 | 0.000 | -0.036 | -0.028 | 0.000 |
| H | 5.376 | -2.532 | 0.00 | -0.264 | 0.033 | -0.000 | 0.101 | 0.014 | 0.000 |
| N | 2.395 | -2.395 | 0.00 | -0.005 | -0.011 | -0.000 | -0.010 | 0.000 | -0.000 |
| C | 0.708 | -6.591 | 0.00 | -0.013 | -0.035 | -0.000 | 0.012 | 0.010 | 0.000 |
| N | -1.988 | -0.000 | 0.00 | -0.012 | 0.162 | -0.000 | -0.047 | -0.041 | 0.000 |
| C | -0.708 | -6.591 | 0.00 | 0.017 | -0.035 | 0.000 | 0.005 | 0.008 | -0.000 |
| N  | -0.000 | -1.988 | 0.00 |  0.041 | -0.047 |  0.000 |  0.162 |  0.012 |  0.000 |
| C  |  1.436 | -5.389 | 0.00 |  0.023 |  0.008 |  0.000 |  0.020 | -0.015 | -0.000 |
| C  | -1.436 | -5.389 | 0.00 | -0.011 |  0.014 | -0.000 |  0.028 |  0.010 |  0.000 |
| C  |  0.708 | -4.188 | 0.00 |  0.025 |  0.078 | -0.000 | -0.046 | -0.010 |  0.000 |
| C  | -0.708 | -4.188 | 0.00 | -0.043 |  0.073 |  0.000 | -0.029 | -0.027 |  0.000 |
| H  |  1.243 | -7.549 | 0.00 | -0.285 | -0.190 |  0.000 |  0.090 |  0.054 | -0.000 |
| H  | -1.243 | -7.549 | 0.00 |  0.293 | -0.193 |  0.000 | -0.056 |  0.043 | -0.000 |
| C  |  1.124 | -2.784 | 0.00 | -0.053 |  0.012 |  0.000 | -0.079 |  0.070 | -0.000 |
| C  | -1.124 | -2.784 | 0.00 |  0.009 | -0.022 | -0.000 | -0.095 | -0.068 | -0.000 |
| H  |  2.532 | -5.376 | 0.00 |  0.028 | -0.036 | -0.000 |  0.022 | -0.280 |  0.000 |
| C  | -2.784 | -1.124 | 0.00 | -0.070 | -0.079 |  0.000 |  0.012 |  0.053 | -0.000 |
| H  | -2.532 | -5.376 | 0.00 | -0.014 |  0.101 | -0.000 |  0.033 |  0.264 | -0.000 |
| C  | -2.784 |  1.124 | 0.00 |  0.068 | -0.095 |  0.000 | -0.022 | -0.009 | -0.000 |
| C  | -4.188 | -0.708 | 0.00 |  0.010 | -0.046 |  0.000 |  0.078 | -0.025 |  0.000 |
| C  | -4.188 |  0.708 | 0.00 |  0.027 | -0.029 | -0.000 |  0.073 |  0.043 | -0.000 |
| C  | -5.389 | -1.436 | 0.00 |  0.015 |  0.020 | -0.000 |  0.008 | -0.023 | -0.000 |
| N  | -2.395 | -2.395 | 0.00 | -0.000 | -0.010 | -0.000 | -0.011 |  0.005 |  0.000 |
| C  | -5.389 |  1.436 | 0.00 | -0.010 |  0.028 | -0.000 |  0.014 |  0.011 |  0.000 |
| N  | -2.395 |  2.395 | 0.00 | -0.005 | -0.011 | -0.000 | -0.010 |  0.000 |  0.000 |
| C  | -6.591 | -0.708 | 0.00 | -0.010 |  0.012 |  0.000 | -0.035 |  0.013 |  0.000 |
| C  | -6.591 |  0.708 | 0.00 | -0.008 |  0.005 | -0.000 | -0.035 | -0.017 | -0.000 |
| H  | -5.376 | -2.532 | 0.00 |  0.280 |  0.022 |  0.000 | -0.036 | -0.028 |  0.000 |
| H  | -5.376 |  2.532 | 0.00 | -0.264 |  0.033 |  0.000 |  0.101 |  0.014 | -0.000 |
| H  | -7.549 | -1.243 | 0.00 | -0.054 |  0.090 | -0.000 | -0.190 |  0.285 | -0.000 |
| H  | -7.549 |  1.243 | 0.00 | -0.043 | -0.056 |  0.000 | -0.193 | -0.293 | -0.000 |
| Cu |  0.000 |  0.000 | 0.00 | -0.000 | -0.000 |  0.000 | -0.000 |  0.000 | -0.000 |

## S7.3   1336 cm$^{-1}$

Table S5: Eigenvectors for the 1336 cm$^{-1}$ eigenmode of CuPc. Each eigenvector is normalized to one and not mass-weighted. $\vec{a}_i$ and $\vec{b}_i$ correspond to the eigenvector components pointing into one of the two planar Cartesian coordinate directions. The positions of the atoms in space are denoted as $x$, $y$, $z$ with units in Angstrom

| Atom | x | y | z | $\vec{a}_x$ | $\vec{a}_y$ | $\vec{a}_z$ | $\vec{b}_x$ | $\vec{b}_y$ | $\vec{b}_z$ |
|---|---|---|---|---|---|---|---|---|---|
| C | 0.708 | 4.188 | 0.00 | -0.111 | 0.069 | -0.000 | 0.082 | -0.044 | -0.000 |
| C | 1.436 | 5.389 | 0.00 | 0.014 | 0.044 | -0.000 | 0.005 | -0.040 | 0.000 |
| C | 0.708 | 6.591 | 0.00 | 0.011 | -0.030 | -0.000 | -0.016 | 0.024 | -0.000 |
| C | -0.708 | 6.591 | 0.00 | -0.013 | -0.016 | 0.000 | 0.015 | 0.035 | -0.000 |
| C | -1.436 | 5.389 | 0.00 | 0.008 | 0.028 | 0.000 | 0.012 | -0.052 | 0.000 |
| C | -0.708 | 4.188 | 0.00 | 0.053 | 0.026 | -0.000 | -0.127 | -0.077 | -0.000 |
| H | 1.243 | 7.549 | 0.00 | 0.048 | -0.052 | -0.000 | -0.032 | 0.035 | 0.000 |
| H | -1.243 | 7.549 | 0.00 | -0.019 | -0.022 | -0.000 | 0.055 | 0.059 | 0.000 |
| H | -2.532 | 5.376 | 0.00 | 0.011 | -0.046 | -0.000 | 0.011 | -0.011 | -0.000 |
| H | 2.532 | 5.376 | 0.00 | 0.014 | 0.022 | 0.000 | 0.008 | 0.042 | 0.000 |
| C | -1.124 | 2.784 | 0.00 | 0.156 | -0.159 | -0.000 | -0.065 | 0.292 | 0.000 |
| N | 0.000 | 1.988 | 0.00 | 0.018 | 0.211 | 0.000 | 0.014 | -0.270 | 0.000 |
| C | 1.124 | 2.784 | 0.00 | -0.025 | -0.245 | 0.000 | 0.167 | 0.224 | -0.000 |
| N | 2.395 | 2.395 | 0.00 | -0.061 | 0.038 | -0.000 | -0.087 | -0.099 | -0.000 |
| C | 6.591 | 0.708 | 0.00 | 0.035 | -0.015 | -0.000 | 0.016 | -0.013 | 0.000 |
| C | 6.591 | -0.708 | 0.00 | 0.024 | 0.016 | -0.000 | 0.030 | 0.011 | 0.000 |
| N | 1.988 | 0.000 | 0.00 | -0.270 | -0.014 | -0.000 | -0.211 | 0.018 | 0.000 |
| C | 5.389 | 1.436 | 0.00 | -0.052 | -0.012 | -0.000 | -0.028 | 0.008 | -0.000 |
| C | 5.389 | -1.436 | 0.00 | -0.040 | -0.005 | 0.000 | -0.044 | 0.014 | -0.000 |
| C | 4.188 | 0.708 | 0.00 | -0.077 | 0.127 | -0.000 | -0.026 | 0.053 | 0.000 |
| C | 4.188 | -0.708 | 0.00 | -0.044 | -0.082 | -0.000 | -0.069 | -0.111 | 0.000 |

| | | | | | | | | |
|---|---|---|---|---|---|---|---|---|
| H | 7.549 | 1.243 | 0.00 | 0.059 | -0.055 | 0.000 | 0.022 | -0.020 | -0.000 |
| H | 7.549 | -1.243 | 0.00 | 0.035 | 0.032 | 0.000 | 0.052 | 0.048 | -0.000 |
| C | 2.784 | 1.124 | 0.00 | 0.292 | 0.065 | 0.000 | 0.159 | 0.156 | -0.000 |
| C | 2.784 | -1.124 | 0.00 | 0.224 | -0.167 | 0.000 | 0.245 | -0.025 | 0.000 |
| H | 5.376 | 2.532 | 0.00 | -0.011 | -0.011 | 0.000 | 0.046 | 0.011 | 0.000 |
| H | 5.376 | -2.532 | 0.00 | 0.042 | -0.008 | 0.000 | -0.022 | 0.014 | 0.000 |
| N | 2.395 | -2.395 | 0.00 | -0.099 | 0.087 | -0.000 | -0.038 | -0.061 | -0.000 |
| C | 0.708 | -6.591 | 0.00 | -0.013 | -0.016 | -0.000 | 0.015 | 0.035 | 0.000 |
| N | -1.988 | -0.000 | 0.00 | -0.270 | -0.014 | 0.000 | -0.211 | 0.018 | 0.000 |
| C | -0.708 | -6.591 | 0.00 | 0.011 | -0.030 | 0.000 | -0.016 | 0.024 | -0.000 |
| N | -0.000 | -1.988 | 0.00 | 0.018 | 0.211 | 0.000 | 0.014 | -0.270 | 0.000 |
| C | 1.436 | -5.389 | 0.00 | 0.008 | 0.028 | 0.000 | 0.012 | -0.052 | -0.000 |
| C | -1.436 | -5.389 | 0.00 | 0.014 | 0.044 | -0.000 | 0.005 | -0.040 | 0.000 |
| C | 0.708 | -4.188 | 0.00 | 0.053 | 0.026 | -0.000 | -0.127 | -0.077 | 0.000 |
| C | -0.708 | -4.188 | 0.00 | -0.111 | 0.069 | 0.000 | 0.082 | -0.044 | 0.000 |
| H | 1.243 | -7.549 | 0.00 | -0.020 | -0.022 | 0.000 | 0.055 | 0.059 | -0.000 |
| H | -1.243 | -7.549 | 0.00 | 0.048 | -0.052 | 0.000 | -0.032 | 0.035 | -0.000 |
| C | 1.124 | -2.784 | 0.00 | 0.156 | -0.159 | -0.000 | -0.065 | 0.292 | -0.000 |
| C | -1.124 | -2.784 | 0.00 | -0.025 | -0.245 | -0.000 | 0.167 | 0.224 | -0.000 |
| H | 2.532 | -5.376 | 0.00 | 0.011 | -0.046 | -0.000 | 0.011 | -0.011 | 0.000 |
| C | -2.784 | -1.124 | 0.00 | 0.292 | 0.065 | 0.000 | 0.159 | 0.156 | -0.000 |
| H | -2.532 | -5.376 | 0.00 | 0.014 | 0.022 | 0.000 | 0.008 | 0.042 | -0.000 |
| C | -2.784 | 1.124 | 0.00 | 0.224 | -0.167 | -0.000 | 0.245 | -0.025 | -0.000 |
| C | -4.188 | -0.708 | 0.00 | -0.077 | 0.127 | 0.000 | -0.026 | 0.053 | 0.000 |
| C | -4.188 | 0.708 | 0.00 | -0.044 | -0.082 | -0.000 | -0.069 | -0.111 | -0.000 |
| C | -5.389 | -1.436 | 0.00 | -0.052 | -0.012 | -0.000 | -0.028 | 0.008 | -0.000 |
| N | -2.395 | -2.395 | 0.00 | -0.061 | 0.038 | -0.000 | -0.087 | -0.099 | 0.000 |

| Atom | x | y | z | $\vec{a}_x$ | $\vec{a}_y$ | $\vec{a}_z$ | $\vec{b}_x$ | $\vec{b}_y$ | $\vec{b}_z$ |
|---|---|---|---|---|---|---|---|---|---|
| C | -5.389 | 1.436 | 0.00 | -0.040 | -0.005 | 0.000 | -0.044 | 0.014 | 0.000 |
| N | -2.395 | 2.395 | 0.00 | -0.099 | 0.087 | -0.000 | -0.038 | -0.061 | 0.000 |
| C | -6.591 | -0.708 | 0.00 | 0.035 | -0.015 | 0.000 | 0.016 | -0.013 | 0.000 |
| C | -6.591 | 0.708 | 0.00 | 0.024 | 0.016 | -0.000 | 0.030 | 0.011 | -0.000 |
| H | -5.376 | -2.532 | 0.00 | -0.011 | -0.011 | -0.000 | 0.046 | 0.011 | 0.000 |
| H | -5.376 | 2.532 | 0.00 | 0.042 | -0.008 | -0.000 | -0.022 | 0.014 | -0.000 |
| H | -7.549 | -1.243 | 0.00 | 0.059 | -0.055 | -0.000 | 0.022 | -0.020 | -0.000 |
| H | -7.549 | 1.243 | 0.00 | 0.035 | 0.032 | 0.000 | 0.052 | 0.048 | -0.000 |
| Cu | 0.000 | 0.000 | 0.00 | 0.004 | -0.003 | 0.000 | 0.003 | 0.004 | -0.000 |

## S7.4  1526 cm$^{-1}$

Table S6: Eigenvectors for the 1526 cm$^{-1}$ eigenmode of CuPc. Each eigenvector is normalized to one and not mass-weighted. $\vec{a}_i$ and $\vec{b}_i$ correspond to the eigenvector components pointing into one of the two planar Cartesian coordinate directions. The positions of the atoms in space are denoted as $x$, $y$, $z$ with units in Angstrom

| Atom | x | y | z | $\vec{a}_x$ | $\vec{a}_y$ | $\vec{a}_z$ | $\vec{b}_x$ | $\vec{b}_y$ | $\vec{b}_z$ |
|---|---|---|---|---|---|---|---|---|---|
| C | 0.708 | 4.188 | 0.00 | -0.027 | -0.024 | -0.000 | 0.052 | -0.024 | -0.000 |
| C | 1.436 | 5.389 | 0.00 | 0.006 | 0.026 | -0.000 | -0.008 | -0.017 | 0.000 |
| C | 0.708 | 6.591 | 0.00 | 0.004 | -0.012 | -0.000 | -0.028 | 0.016 | -0.000 |
| C | -0.708 | 6.591 | 0.00 | -0.004 | 0.008 | 0.000 | 0.028 | 0.018 | -0.000 |
| C | -1.436 | 5.389 | 0.00 | 0.003 | -0.020 | 0.000 | 0.010 | -0.023 | 0.000 |
| C | -0.708 | 4.188 | 0.00 | -0.012 | 0.029 | -0.000 | -0.057 | -0.016 | -0.000 |
| H | 1.243 | 7.549 | 0.00 | -0.006 | -0.008 | -0.000 | -0.006 | 0.002 | 0.000 |
| H | -1.243 | 7.549 | 0.00 | -0.007 | 0.007 | -0.000 | 0.004 | 0.004 | 0.000 |
| H | -2.532 | 5.376 | 0.00 | 0.004 | 0.056 | -0.000 | 0.013 | -0.024 | -0.000 |
| H | 2.532 | 5.376 | 0.00 | 0.008 | -0.047 | 0.000 | -0.011 | -0.039 | 0.000 |
| C | -1.124 | 2.784 | 0.00 | -0.038 | -0.074 | -0.000 | 0.387 | 0.100 | 0.000 |
| N | 0.000 | 1.988 | 0.00 | -0.101 | 0.001 | 0.000 | -0.015 | -0.006 | 0.000 |
| C | 1.124 | 2.784 | 0.00 | 0.073 | 0.042 | 0.000 | -0.382 | 0.117 | -0.000 |
| N | 2.395 | 2.395 | 0.00 | 0.087 | -0.211 | -0.000 | 0.245 | -0.153 | -0.000 |
| C | 6.591 | 0.708 | 0.00 | -0.018 | 0.028 | -0.000 | 0.008 | 0.004 | 0.000 |
| C | 6.591 | -0.708 | 0.00 | -0.016 | -0.028 | -0.000 | -0.012 | -0.004 | 0.000 |
| N | 1.988 | 0.000 | 0.00 | 0.006 | -0.015 | -0.000 | 0.001 | 0.101 | 0.000 |
| C | 5.389 | 1.436 | 0.00 | 0.024 | 0.010 | -0.000 | -0.020 | -0.003 | -0.000 |
| C | 5.389 | -1.436 | 0.00 | 0.017 | -0.008 | 0.000 | 0.026 | -0.006 | -0.000 |
| C | 4.188 | 0.708 | 0.00 | 0.016 | -0.057 | -0.000 | 0.029 | 0.012 | 0.000 |
| C | 4.188 | -0.708 | 0.00 | 0.024 | 0.052 | -0.000 | -0.024 | 0.027 | 0.000 |
| H | 7.549 | 1.243 | 0.00 | -0.004 | 0.004 | 0.000 | 0.007 | 0.007 | -0.000 |
| H | 7.549 | -1.243 | 0.00 | -0.002 | -0.006 | 0.000 | -0.008 | 0.006 | -0.000 |
| C | 2.784 | 1.124 | 0.00 | -0.100 | 0.387 | 0.000 | -0.073 | 0.038 | -0.000 |
| C | 2.784 | -1.124 | 0.00 | -0.117 | -0.382 | 0.000 | 0.042 | -0.073 | 0.000 |
| H | 5.376 | 2.532 | 0.00 | 0.024 | 0.013 | 0.000 | 0.056 | -0.004 | 0.000 |
| H | 5.376 | -2.532 | 0.00 | 0.039 | -0.011 | 0.000 | -0.047 | -0.008 | 0.000 |
| N | 2.395 | -2.395 | 0.00 | 0.153 | 0.245 | -0.000 | -0.211 | -0.087 | -0.000 |
| C | 0.708 | -6.591 | 0.00 | -0.004 | 0.008 | -0.000 | 0.028 | 0.018 | 0.000 |
| N | -1.988 | -0.000 | 0.00 | 0.006 | -0.015 | 0.000 | 0.001 | 0.101 | 0.000 |
| C | -0.708 | -6.591 | 0.00 | 0.004 | -0.012 | 0.000 | -0.028 | 0.016 | -0.000 |
| N | -0.000 | -1.988 | 0.00 | -0.101 | 0.001 | 0.000 | -0.015 | -0.006 | 0.000 |
| C | 1.436 | -5.389 | 0.00 | 0.003 | -0.020 | 0.000 | 0.010 | -0.024 | -0.000 |
| C | -1.436 | -5.389 | 0.00 | 0.006 | 0.026 | -0.000 | -0.008 | -0.017 | 0.000 |
| C | 0.708 | -4.188 | 0.00 | -0.012 | 0.029 | -0.000 | -0.057 | -0.016 | 0.000 |
| C | -0.708 | -4.188 | 0.00 | -0.027 | -0.024 | 0.000 | 0.052 | -0.024 | 0.000 |
| H | 1.243 | -7.549 | 0.00 | -0.007 | 0.007 | 0.000 | 0.004 | 0.004 | -0.000 |
| H | -1.243 | -7.549 | 0.00 | -0.006 | -0.008 | 0.000 | -0.006 | 0.002 | -0.000 |
| C | 1.124 | -2.784 | 0.00 | -0.038 | -0.074 | -0.000 | 0.387 | 0.100 | -0.000 |
| C | -1.124 | -2.784 | 0.00 | 0.073 | 0.042 | -0.000 | -0.381 | 0.117 | -0.000 |
| H | 2.532 | -5.376 | 0.00 | 0.004 | 0.056 | -0.000 | 0.013 | -0.024 | 0.000 |
| C | -2.784 | -1.124 | 0.00 | -0.100 | 0.387 | 0.000 | -0.074 | 0.038 | -0.000 |
| H | -2.532 | -5.376 | 0.00 | 0.008 | -0.047 | 0.000 | -0.011 | -0.039 | -0.000 |
| C | -2.784 | 1.124 | 0.00 | -0.117 | -0.382 | -0.000 | 0.042 | -0.073 | -0.000 |
| C | -4.188 | -0.708 | 0.00 | 0.016 | -0.057 | 0.000 | 0.030 | 0.012 | 0.000 |
| C | -4.188 | 0.708 | 0.00 | 0.024 | 0.052 | -0.000 | -0.024 | 0.027 | -0.000 |
| C | -5.389 | -1.436 | 0.00 | 0.024 | 0.010 | -0.000 | -0.020 | -0.003 | -0.000 |
| N | -2.395 | -2.395 | 0.00 | 0.087 | -0.211 | -0.000 | 0.245 | -0.153 | 0.000 |
| C | -5.389 | 1.436 | 0.00 | 0.017 | -0.008 | 0.000 | 0.026 | -0.006 | 0.000 |
| N | -2.395 | 2.395 | 0.00 | 0.153 | 0.245 | -0.000 | -0.211 | -0.087 | 0.000 |
| C | -6.591 | -0.708 | 0.00 | -0.018 | 0.028 | 0.000 | 0.008 | 0.004 | 0.000 |
| C | -6.591 | 0.708 | 0.00 | -0.016 | -0.028 | -0.000 | -0.012 | -0.004 | -0.000 |
| H | -5.376 | -2.532 | 0.00 | 0.024 | 0.013 | -0.000 | 0.056 | -0.004 | 0.000 |
| H | -5.376 | 2.532 | 0.00 | 0.039 | -0.011 | -0.000 | -0.047 | -0.008 | -0.000 |
| H | -7.549 | -1.243 | 0.00 | -0.004 | 0.004 | -0.000 | 0.007 | 0.007 | -0.000 |
| H | -7.549 | 1.243 | 0.00 | -0.002 | -0.006 | 0.000 | -0.008 | 0.006 | -0.000 |
| Cu | 0.000 | 0.000 | 0.00 | -0.003 | -0.001 | 0.000 | -0.001 | 0.003 | -0.000 |

## S7.5  3144 cm$^{-1}$

Table S7: Eigenvectors for the 3144 cm$^{-1}$ eigenmode of CuPc. Each eigenvector is normalized to one and not mass-weighted. $\vec{a}_i$ and $\vec{b}_i$ correspond to the eigenvector components pointing into one of the two planar Cartesian coordinate directions. The positions of the atoms in space are denoted as $x$, $y$, $z$ with units in Angstrom

| Atom | x | y | z | $\vec{a}_x$ | $\vec{a}_y$ | $\vec{a}_z$ | $\vec{b}_x$ | $\vec{b}_y$ | $\vec{b}_z$ |
|---|---|---|---|---|---|---|---|---|---|
| C | 0.708 | 4.188 | 0.00 | -0.000 | -0.002 | -0.000 | -0.000 | -0.000 | -0.000 |
| C | 1.436 | 5.389 | 0.00 | 0.039 | -0.001 | -0.000 | 0.011 | -0.000 | 0.000 |
| C | 0.708 | 6.591 | 0.00 | 0.009 | 0.018 | -0.000 | 0.002 | 0.005 | -0.000 |
| C | -0.708 | 6.591 | 0.00 | -0.009 | 0.018 | 0.000 | -0.002 | 0.004 | -0.000 |
| C | -1.436 | 5.389 | 0.00 | -0.040 | -0.001 | 0.000 | -0.008 | -0.000 | 0.000 |
| C | -0.708 | 4.188 | 0.00 | 0.000 | -0.002 | -0.000 | 0.000 | -0.000 | -0.000 |
| H | 1.243 | 7.549 | 0.00 | -0.105 | -0.189 | -0.000 | -0.027 | -0.049 | 0.000 |
| H | -1.243 | 7.549 | 0.00 | 0.106 | -0.191 | -0.000 | 0.023 | -0.041 | 0.000 |
| H | -2.532 | 5.376 | 0.00 | 0.449 | 0.003 | -0.000 | 0.085 | 0.001 | -0.000 |
| H | 2.532 | 5.376 | 0.00 | -0.440 | 0.003 | 0.000 | -0.125 | 0.001 | 0.000 |
| C | -1.124 | 2.784 | 0.00 | -0.000 | -0.000 | -0.000 | -0.000 | -0.000 | 0.000 |
| N | 0.000 | 1.988 | 0.00 | -0.000 | 0.000 | 0.000 | 0.000 | 0.000 | 0.000 |
| C | 1.124 | 2.784 | 0.00 | 0.000 | -0.000 | 0.000 | 0.000 | -0.000 | -0.000 |
| N | 2.395 | 2.395 | 0.00 | -0.000 | -0.000 | -0.000 | -0.000 | 0.000 | -0.000 |
| C | 6.591 | 0.708 | 0.00 | 0.004 | 0.002 | -0.000 | -0.018 | -0.009 | 0.000 |
| C | 6.591 | -0.708 | 0.00 | 0.005 | -0.002 | -0.000 | -0.018 | 0.009 | 0.000 |
| N | 1.988 | 0.000 | 0.00 | 0.000 | -0.000 | -0.000 | -0.000 | -0.000 | 0.000 |
| C | 5.389 | 1.436 | 0.00 | -0.000 | 0.008 | -0.000 | 0.001 | -0.039 | -0.000 |
| C | 5.389 | -1.436 | 0.00 | -0.000 | -0.011 | 0.000 | 0.001 | 0.039 | -0.000 |
| C | 4.188 | 0.708 | 0.00 | -0.000 | -0.000 | -0.000 | 0.002 | 0.000 | 0.000 |
| C | 4.188 | -0.708 | 0.00 | -0.000 | 0.000 | -0.000 | 0.002 | -0.000 | 0.000 |
| H | 7.549 | 1.243 | 0.00 | -0.041 | -0.023 | 0.000 | 0.187 | 0.104 | -0.000 |
| H | 7.549 | -1.243 | 0.00 | -0.049 | 0.027 | 0.000 | 0.185 | -0.103 | -0.000 |
| C | 2.784 | 1.124 | 0.00 | -0.000 | 0.000 | 0.000 | 0.000 | -0.000 | -0.000 |
| C | 2.784 | -1.124 | 0.00 | -0.000 | -0.000 | 0.000 | 0.000 | 0.000 | 0.000 |
| H | 5.376 | 2.532 | 0.00 | 0.001 | -0.085 | 0.000 | -0.003 | 0.439 | 0.000 |
| H | 5.376 | -2.532 | 0.00 | 0.001 | 0.125 | 0.000 | -0.003 | -0.430 | 0.000 |
| N | 2.395 | -2.395 | 0.00 | 0.000 | 0.000 | -0.000 | 0.000 | -0.000 | -0.000 |

| | | | | | | | | |
|---|---|---|---|---|---|---|---|---|
| C | 0.708 | -6.591 | 0.00 | -0.009 | 0.018 | -0.000 | -0.002 | 0.004 | 0.000 |
| N | -1.988 | -0.000 | 0.00 | 0.000 | -0.000 | 0.000 | -0.000 | -0.000 | 0.000 |
| C | -0.708 | -6.591 | 0.00 | 0.009 | 0.017 | 0.000 | 0.002 | 0.005 | -0.000 |
| N | -0.000 | -1.988 | 0.00 | -0.000 | 0.000 | 0.000 | 0.000 | 0.000 | 0.000 |
| C | 1.436 | -5.389 | 0.00 | -0.039 | -0.001 | 0.000 | -0.007 | -0.000 | -0.000 |
| C | -1.436 | -5.389 | 0.00 | 0.038 | -0.001 | -0.000 | 0.011 | -0.000 | 0.000 |
| C | 0.708 | -4.188 | 0.00 | 0.000 | -0.002 | -0.000 | 0.000 | -0.000 | 0.000 |
| C | -0.708 | -4.188 | 0.00 | -0.000 | -0.002 | 0.000 | -0.000 | -0.000 | 0.000 |
| H | 1.243 | -7.549 | 0.00 | 0.102 | -0.183 | 0.000 | 0.022 | -0.039 | -0.000 |
| H | -1.243 | -7.549 | 0.00 | -0.101 | -0.181 | 0.000 | -0.026 | -0.047 | -0.000 |
| C | 1.124 | -2.784 | 0.00 | -0.000 | -0.000 | -0.000 | -0.000 | -0.000 | -0.000 |
| C | -1.124 | -2.784 | 0.00 | 0.000 | -0.000 | 0.000 | 0.000 | -0.000 | -0.000 |
| H | 2.532 | -5.376 | 0.00 | 0.429 | 0.003 | -0.000 | 0.080 | 0.001 | 0.000 |
| C | -2.784 | -1.124 | 0.00 | -0.000 | 0.000 | 0.000 | 0.000 | -0.000 | -0.000 |
| H | -2.532 | -5.376 | 0.00 | -0.420 | 0.003 | 0.000 | -0.121 | 0.001 | -0.000 |
| C | -2.784 | 1.124 | 0.00 | -0.000 | -0.000 | -0.000 | 0.000 | 0.000 | -0.000 |
| C | -4.188 | -0.708 | 0.00 | -0.000 | -0.000 | 0.000 | 0.002 | 0.000 | 0.000 |
| C | -4.188 | 0.708 | 0.00 | -0.000 | 0.000 | -0.000 | 0.002 | -0.000 | -0.000 |
| C | -5.389 | -1.436 | 0.00 | -0.000 | 0.007 | -0.000 | 0.001 | -0.039 | -0.000 |
| N | -2.395 | -2.395 | 0.00 | -0.000 | -0.000 | -0.000 | -0.000 | 0.000 | 0.000 |
| C | -5.389 | 1.436 | 0.00 | -0.000 | -0.011 | 0.000 | 0.001 | 0.039 | 0.000 |
| N | -2.395 | 2.395 | 0.00 | 0.000 | 0.000 | -0.000 | 0.000 | -0.000 | 0.000 |
| C | -6.591 | -0.708 | 0.00 | 0.004 | 0.002 | 0.000 | -0.018 | -0.009 | 0.000 |
| C | -6.591 | 0.708 | 0.00 | 0.005 | -0.002 | -0.000 | -0.018 | 0.009 | -0.000 |
| H | -5.376 | -2.532 | 0.00 | 0.001 | -0.080 | -0.000 | -0.003 | 0.440 | 0.000 |
| H | -5.376 | 2.532 | 0.00 | 0.001 | 0.121 | -0.000 | -0.003 | -0.430 | -0.000 |
| H | -7.549 | -1.243 | 0.00 | -0.039 | -0.022 | -0.000 | 0.187 | 0.104 | -0.000 |
| H | -7.549 | 1.243 | 0.00 | -0.047 | 0.026 | 0.000 | 0.185 | -0.103 | -0.000 |
| Cu | 0.000 | 0.000 | 0.00 | -0.000 | -0.000 | 0.000 | 0.000 | -0.000 | -0.000 |

# References

(1) Bannwarth, C.; Ehlert, S.; Grimme, S. GFN2-xTB—An Accurate and Broadly Parametrized Self-Consistent Tight-Binding Quantum Chemical Method with Multipole Electrostatics and Density-Dependent Dispersion Contributions. *Journal of Chemical Theory and Computation* **2019**, *15*, 1652–1671.

(2) Bannwarth, C.; Caldeweyher, E.; Ehlert, S.; Hansen, A.; Pracht, P.; Seibert, J.; Spicher, S.; Grimme, S. Extended tight-binding quantum chemistry methods. *WIREs Computational Molecular Science* **2021**, *11*, e1493.

(3) Epifanovsky, E.; et al. Software for the frontiers of quantum chemistry: An overview of developments in the Q-Chem 5 package. *The Journal of Chemical Physics* **2021**, *155*, 084801.

(4) Darling, B. T.; Dennison, D. M. The Water Vapor Molecule. *Phys. Rev.* **1940**, *57*, 128–139.

(5) Watson, J. K. Simplification of the molecular vibration-rotation hamiltonian. *Molecular Physics* **1968**, *15*, 479–490.

(6) Moss, R.; Perry, A. The vibrational Zeeman effect. *Molecular Physics* **1973**, *25*, 1121–1134.

(7) Gauss, J.; Ruud, K.; Helgaker, T. Perturbation-dependent atomic orbitals for the calculation of spin-rotation constants and rotational g tensors. *The Journal of Chemical Physics* **1996**, *105*, 2804–2812.

(8) Wang, B.; Tam, C. N.; Keiderling, T. A. Vibrational Zeeman effect for the $\nu_4$ mode of haloforms (HC$X_3$) determined by magnetic vibrational circular dichroism. *Phys. Rev. Lett.* **1993**, *71*, 979–982.

(9) Wang, B.; Keiderling, T. A. Measurement of the vibrational Zeeman effect for HCF3 using magnetic vibrational circular dichroism. *The Journal of Chemical Physics* **1994**, *101*, 905–911.

(10) Dauxois, T.; Peyrard, M. *Physics of Solitons*; Cambridge University Press, 2006.